\documentclass{article}
\usepackage{amsmath,amsfonts}
\usepackage[left=1in, right=1in, top=1in, bottom=1in]{geometry}
\usepackage{algorithmic}
\usepackage{algorithm}
\usepackage{enumitem}
\usepackage{array}
\usepackage{textcomp}
\usepackage{stfloats}
\usepackage{url}
\usepackage{verbatim}
\usepackage{graphicx}
\usepackage{caption}
\usepackage{subcaption}
\usepackage{cite}
\usepackage{soul,color}
\usepackage{multirow} 
\usepackage{amsmath}
\usepackage{adjustbox}
\usepackage{array}
\usepackage[table]{xcolor}
\usepackage{tikz}
\usepackage[most]{tcolorbox}
\usepackage{pifont}
\usepackage{longtable}
\usepackage{graphicx}
\usepackage{makecell}
\usepackage{authblk}
\usepackage{pdflscape}

\newcommand{\cmark}{\ding{51}} 
\newcommand{\xmark}{\ding{55}} 

\title{Robust ML-based Detection of Conventional, LLM-Generated, and Adversarial Phishing Emails Using Advanced Text Preprocessing}
\author[1]{Deeksha Hareesha Kulal}
\author[1]{Chidozie Princewill Arannonu}
\author[2]{Afsah Anwar}
\author[3]{Nidhi Rastogi}
\author[1]{Quamar Niyaz\thanks{Corresponding author: Quamar Niyaz (qniyaz@pnw.edu)}}
\affil[1]{Department of Electrical and Computer Engineering, Purdue University Northwest, Hammond, IN 46323, Email: \{dhareesh, carannon, qniyaz\}@pnw.edu}
\affil[2]{Department of Computer Science, University of New Mexico, Albuquerque, NM 87131, Email: afsah@unm.edu}
\affil[3]{Department of Software Engineering, Rochester Institute of Technology, Rochester, NY 14623, Email: nidhi.rastogi@rit.edu}

\date{}
\begin{document}

\maketitle

\begin{abstract}
Phishing remains a critical cybersecurity threat, especially with the advent of large language models (LLMs) capable of generating highly convincing malicious content. Unlike earlier phishing attempts which are identifiable by grammatical errors, misspellings, incorrect phrasing, and inconsistent formatting, LLM generated emails are grammatically sound, contextually relevant, and linguistically natural. These advancements make phishing emails increasingly difficult to distinguish from legitimate ones, challenging traditional detection mechanisms. Conventional phishing detection systems often fail when faced with emails crafted by LLMs or manipulated using adversarial perturbation techniques. To address this challenge, we propose a robust phishing email detection system featuring an enhanced text preprocessing pipeline. This pipeline includes spelling correction and word splitting to counteract adversarial modifications and improve detection accuracy. Our approach integrates widely adopted natural language processing (NLP) feature extraction techniques and machine learning algorithms. We evaluate our models on publicly available datasets comprising both phishing and legitimate emails, achieving a detection accuracy of 94.26\% and F1-score of 84.39\% in model deployment setting. To assess robustness, we further evaluate our models using adversarial phishing samples generated by four attack methods in Python TextAttack framework. Additionally, we evaluate models' performance against phishing emails generated by LLMs including ChatGPT and Llama. Results highlight the resilience of models against evolving AI-powered phishing threats. 
\end{abstract}


\section{Introduction}
Phishing is a cyberattack technique employed by malicious actors to obtain sensitive information from individuals, such as login credentials, personal identification details, and financial data. Beyond compromising user privacy and system security, phishing can also lead to significant economic losses. Common phishing methods include deceptive emails, fraudulent job offers, and phone scams. The use of artificial intelligence (AI) in crafting sophisticated phishing contents has further increased the threat. Despite increased awareness about phishing, individuals continue to fall victim to these scams. According to Gallup data, approximately 15\% of U.S. adults reported that they or one of their family members were victims of a scam in the past year~\cite{gallup2024survey}. While phishing targets users across various demographics, those aged 35-44 have been identified as particularly vulnerable, often with the highest financial losses~\cite{gallup2024demo}. The Internet Crime Complaint Center (IC3) 2018 Internet Crime Report documented approximately \$2.7 billion in losses based on 351,937 reported incidents of suspected internet-related crimes~\cite{economicimpact2024}. Phishing attacks not only harm individuals but also impose regulatory and reputational consequences on affected institutions. In addition to financial impact, phishing can also negatively impact victims' mental well-being and a reduced sense of digital safety.  

Prior to large language models (LLMs) and generative AI tools, phishing emails were often easy to identify due to poor grammar, spelling errors, unusual word combinations, and inconsistent formatting. These characteristics, combined with suspicious links or attachments, served as clear indicators of malicious intent. However, with these AI tools, phishing emails have become significantly more sophisticated and harder to detect. Attackers have created specialized LLMs like FraudGPT \cite{fraudgpt} and WormGPT \cite{wormgpt} for generating phishing and malicious contents that appear grammatically correct, contextually relevant, linguistically natural, and legitimate. As a result, recipients are more likely to trust the content and act on the instructions, thereby increasing the effectiveness of phishing attacks. This highlights the need for a reevaluation of traditional anti-phishing strategies. Detection mechanisms must now analyze deeper contextual understanding and potential indicators of LLM-generated content, rather than relying solely on surface-level features. In addition, the emerging LLMs and generative AI tools have made phishing attacks more scalable as adversaries can use them to create phishing messages at large volumes through automation~\cite{llmimpact2024,llmsocialengineering2024}. 

As many email services employ AI and machine learning (ML) based phishing detection systems, adversarial attacks on ML models can exploit vulnerabilities in them by manipulating the phishing email contents to evade classification. These changes are often subtle but can significantly degrade the detection capabilities of phishing emails. Understanding and identifying adversarial behavior are essential in today's phishing landscape. Phishing detectors can be attacked via various components like text-based content where text is exploited through subtle changes in subject lines, email body, and embedded URLs. URL-based attacks involve typo-squatting, deceptive subdomains, or URL encoding to bypass filters. In email header spoofing, sender identity is manipulated and exploited. There can be use of misleading images or embedded text to convey malicious intent.

To detect the phishing emails in this advanced AI era and improve the adversarial robustness of phishing detection systems, we implement a robust ML-based phishing email detection system by incorporating advanced text preprocessing techniques namely, spelling correction and word splitting. The following research questions guide our investigation:
\begin{enumerate}
\item \textbf{RQ1}: To what extent can an ML-based phishing detection model, enhanced with spelling correction and word splitting, effectively identify phishing emails?
\item \textbf{RQ2}: Can the proposed phishing detection model maintain high detection accuracy in the presence of adversarially manipulated phishing emails?
\item \textbf{RQ3}: How effectively does the proposed model detect phishing emails generated by LLMs?
\end{enumerate}

\noindent Toward this, we make the following contributions in this work:
\begin{enumerate}
    \item We implement an ML-based phishing email detection system by introducing spelling correction and word splitting addition to traditional text preprocessing techniques. 
    \item We exploit widely adopted feature extraction techniques in natural language processing (NLP) and ML algorithms for the proposed phishing detection system. 
    \item To ensure our models can perform well in real-world scenarios, we use phishing and legitimate emails collected from different time periods for both model development and deployment.
    \item To assess the robustness of the phishing detection models, we simulate various adversarial attack scenarios using Python \texttt{TextAttack} library \cite{textattack}.
    \item We evaluate our models against phishing emails generated via prompt engineering from LLMs including ChatGPT \cite{chatgpt} and Llama \cite{llama}. 
\end{enumerate}

This work significantly extends our previous research \cite{kulal}, which introduced an ML-based phishing detection model incorporating advanced text preprocessing techniques. In the current study, we present a comprehensive evaluation of the proposed models, including comparative analyses with baseline models that utilize only conventional text preprocessing. Furthermore, we assess the robustness of the detection models against the emails that are adversarially crafted and generated by LLMs.

The remainder of this paper is organized as follows. Section \ref{sec:related} reviews prior work on ML-based phishing detection, the application of NLP techniques, and challenges introduced by adversarial manipulations and LLM-generated threats. Section \ref{sec:phishing_detection_system} presents our proposed approach for building and evaluating phishing detection models enhanced with spelling correction and word splitting mechanisms. In Section \ref{sec:adversarial_analysis}, we discuss the implementation of various adversarial attacks aimed at assessing the robustness of the proposed approach. Section \ref{sec:llm_based_phishing} evaluates the models' effectiveness in detecting phishing emails crafted using LLMs. Finally, Section \ref{sec:conclusion} concludes the paper and outlines directions for future research.

\section{Related Work}
\label{sec:related}
Phishing detection using ML has been extensively studied in recent years, with researchers exploring various feature engineering, NLP, and classification techniques. This section summarizes existing contributions in three major categories relevant to our research including ML-based phishing detection models, the role of grammar and linguistic errors, and adversarial attacks on phishing detection models.
\subsection{ML-based Phishing Detection Models}
Early phishing detection models focused on extracting basic textual or structural features from email headers and bodies. However, recent advancements make use of the NLP embeddings such as TF-IDF, Word2Vec, Doc2Vec, and GloVe to capture semantic information from email content. Unnithan et al. performed phishing classification using Doc2Vec and GloVe for NLP feature extraction, combined with classifiers such as SVM, Logistic Regression, and Decision Trees~\cite{unnithan2018phishing}. The best performance was achieved with a Doc2Vec and SVM combination with an accuracy of 88.4\%. Harikrishnan et al. used TF-IDF embeddings with dimensionality reduction techniques such as Singular Value Decomposition (SVD) and Non-Negative Matrix Factorization (NMF) followed by classification~\cite{harikrishnan2018machine}. The best performance was achieved using SVM with TF-IDF and NMF, reaching an F1-score of 94.4\% on the test set. This approach effectively detects phishing emails despite dataset imbalance. Bountakas et al. compared NLP-based feature extraction methods like TF-IDF, Word2Vec, and GloVe with classifiers such as Naive Bayes, Gradient Boosting, and RF~\cite{bountakas2021comparison}. Word2Vec with RF achieved the best results across datasets like Enron and Nazario with an accuracy of 98.95\% and F1-score of 98.97\%. Bountakas and Xenakis~\cite{bountakas2023helphed} proposed HELPHED, a hybrid ensemble learning technique using stacking and soft voting, which integrated lexical, structural, and semantic features. Their ensemble model outperformed standalone classifiers, especially on imbalanced datasets. This paper proposed two ensemble methods named Stacking and Soft Voting that process hybrid features separately using two ML models in parallel with Decision Tree for content, KNN for text. The Soft Voting method achieved the highest detection accuracy of 99.42\% and F1-score of 99.42\% with low training time. The study by Omari evaluates seven ML models for phishing domain detection using the UCI dataset~\cite{omari2023comparative}. Among them, Gradient Boosting and Random Forest models achieved the best performance, with an accuracy of 97.2\% and F1-score of 96.9\%. These models demonstrate strong potential in effectively identifying phishing threats. Somesha and Pais introduced DeepPhishNet, a deep neural network using FastText and Word2Vec embeddings for phishing detection~\cite{somesha2024deepphishnet}. Their model achieved state-of-the-art performance on multiple benchmark datasets. The best one was DNN model with FastText-SkipGram which achieved an accuracy of 99.52\%.

\subsection{Use of Grammar and Word Errors in Phishing Detection}
Several studies have explored the role of grammar and syntactic irregularities in phishing detection. In \cite{sentence_knn_2022}, the authors proposed a sentence-level analysis model leveraging K-Nearest Neighbors (KNN), which focused on identifying grammatical errors, wrong syntax, and inconsistent language usage—common traits found in phishing emails. Their approach demonstrated that sentence structure and linguistic anomalies could serve as useful features for detecting phishing attempts. KNN model's effectiveness relies on key performance metrics with an accuracy of 97\%. Similarly, Park and Taylor analyzed the syntactic structure of sentences, particularly the subjects and objects of verbs, to assess their discriminative power in distinguishing phishing from legitimate emails in \cite{syntactic_features_2015}. They conducted two experiments, one on syntactic similarity and another on verb subject-object comparison and found that while these features were useful for certain verbs, additional research was needed to generalize the findings. These studies indicate syntactic features alone are insufficient for reliable phishing detection, but verb objects offer better discrimination, suggesting future work should focus on semantic analysis for improved robustness. Our approach bears resemblance to the work of Gong et al. \cite{gong2019context}, who introduced a context-aware spelling correction technique aimed at improving spam and toxicity detection. In contrast, our method extends beyond spelling correction by incorporating a word splitting mechanism to effectively handle cases where phishers concatenate words to bypass detection systems. Additionally, our study utilizes a more diverse and substantially larger dataset compared to their work.

\subsection{Adversarial Attacks on Phishing Detection Models}
Recent studies have highlighted the vulnerabilities of spam and phishing detection models to adversarial attacks. In \cite{feature_perturbation_2022}, authors developed methods that can translate adversarial perturbations in the NLP feature embeddings back to a set of malicious words, in the corresponding text, which can cause desirable spam misclassifications by employing Projected Gradient Descent (PGD) attack. The adversarial spam samples were able to bypass spam filters with a success rate exceeding 90\% on ML classifiers. Similarly, the work in \cite{agwep_2023} introduced the Adaptive Gradient-based Word Embedding Perturbations (AG-WEP) framework, which modified spam content using benign synonyms. The resulting adversarial samples reduced the detection accuracy of deep learning spam filters by over 15\%, showing that even gradient-aware models are susceptible to subtle text perturbations. The study in \cite{robustness_phishing_2023} proposed adversarial training using an augmented dataset. The F1-score of a BERT-based phishing classifier improved from 86.2\% to 91.7\% post-adversarial training, demonstrating enhanced robustness. In \cite{bert_adversarial_2023}, authors showed that simple word substitutions and insertions led up to a 30\% drop in classification confidence of BERT spam classifiers, without significantly altering semantic meaning. 

Furthermore, the emergence of generative LLMs has introduced new threats. In \cite{phishbots_llms_2023}, phishing emails generated using ChatGPT, Claude, and Bard were shown to evade detection systems in over 70\% of test cases. The authors developed a BERT-based classifier capable of detecting LLM generated phishing prompts with a precision of 94.5\%. Finally, the authors in \cite{chatgpt_spam_2023} emphasized the potential misuse of ChatGPT in crafting contextually rich, grammar-perfect spam emails. Their experiments revealed that spam detection accuracy dropped from 96\% to 78\% when evaluated on ChatGPT-generated spam, underscoring the challenge these models pose to traditional spam filters.

Together, these works stress the importance of building robust phishing and spam detection systems that can withstand both feature-level perturbations and sophisticated content generated by LLMs. Table \ref{tab:comparison} provides comparison of our work with the above-mentioned existing works.

\section{ML-based Phishing Detection System}
\label{sec:phishing_detection_system}
In this section, we describe the implementation of ML-based phishing detection system. We detail how spelling correction and word splitting are integrated into preprocessing, along with the application of NLP feature extraction methods. We briefly discuss various ML algorithms employed for model development and metrics used for model evaluation. Following that we discuss the performance of the ML models.
\subsection{Dataset Collection and Preprocessing}
We used Millersmiles \cite{millersmile}, Nazario \cite{Nazario}, and Enron \cite{Enron} public datasets to develop and evaluate our ML based phishing email detection system. Millersmiles and Nazario datasets represent phishing emails, while the Enron dataset comprises legitimate emails. Millersmiles dataset was collected by scraping phishing emails from the Millersmiles archive, which maintains a chronological archive of phishing emails dating back to the early 2000. The archive is unstructured, required web scraping for structured extraction. Python's \texttt{requests} library was used to fetch page content, while \texttt{BeautifulSoup} was employed to parse HTML. Nazario corpus consists of 1,564 phishing emails collected from 2015 to 2022. Enron dataset was obtained from Kaggle for legitimate emails. It has approximately 500,000 internal emails exchanged by Enron Corporation employees. 

To ensure the usefulness of our models, we used phishing and legitimate emails collected from different time periods for model development and deployment. For model development, we used 29,500 emails from Millersmiles dataset dated between 2005 and 2014. For evaluating the model in a deployment setting, we used a separate set of 2,927 emails from 2015 to 2024. Phishing emails from Nazario dataset was exclusively used for evaluating models' robustness during model deployment. From Enron dataset, we used 120,000 legitimate emails from 1997 to mid-2001 for model development, and another 22,458 emails from 2001 to 2002 for deployment evaluation. During model development, the data was split into training and test sets using an 80:20 ratio. Specifically, 96,000 legitimate and 24,000 phishing emails were used for training, while 24,000 legitimate and 5,500 phishing emails were used for testing during model development. For the deployment evaluation, we used 22,458 legitimate emails and 4,491 phishing emails. Table \ref{tab:email_distribution} shows the distribution of emails used for model development and deployment phases. The imbalanced distribution of records between legitimate and phishing emails reflects real-world scenarios, where users typically receive a higher volume of legitimate emails compared to phishing attempts.
\begin{table}
\centering
\caption{Emails used for model development and deployment\cite{kulal}}
\label{tab:email_distribution}
\begin{tabular}{|l|c|c|}
\hline
\textbf{} & \textbf{Legitimate Emails} & \textbf{Phishing Emails} \\
\hline
\textbf{Model Development} & & \\
Training dataset & 96,000 & 24,000 \\
Test dataset & 24,000 & 5,500 \\
\hline
\textbf{Model Deployment} & 22,458 & 4,491 \\
\hline
\end{tabular}
\end{table}

\begin{figure*}
\centering
\includegraphics[width=0.9\textwidth]{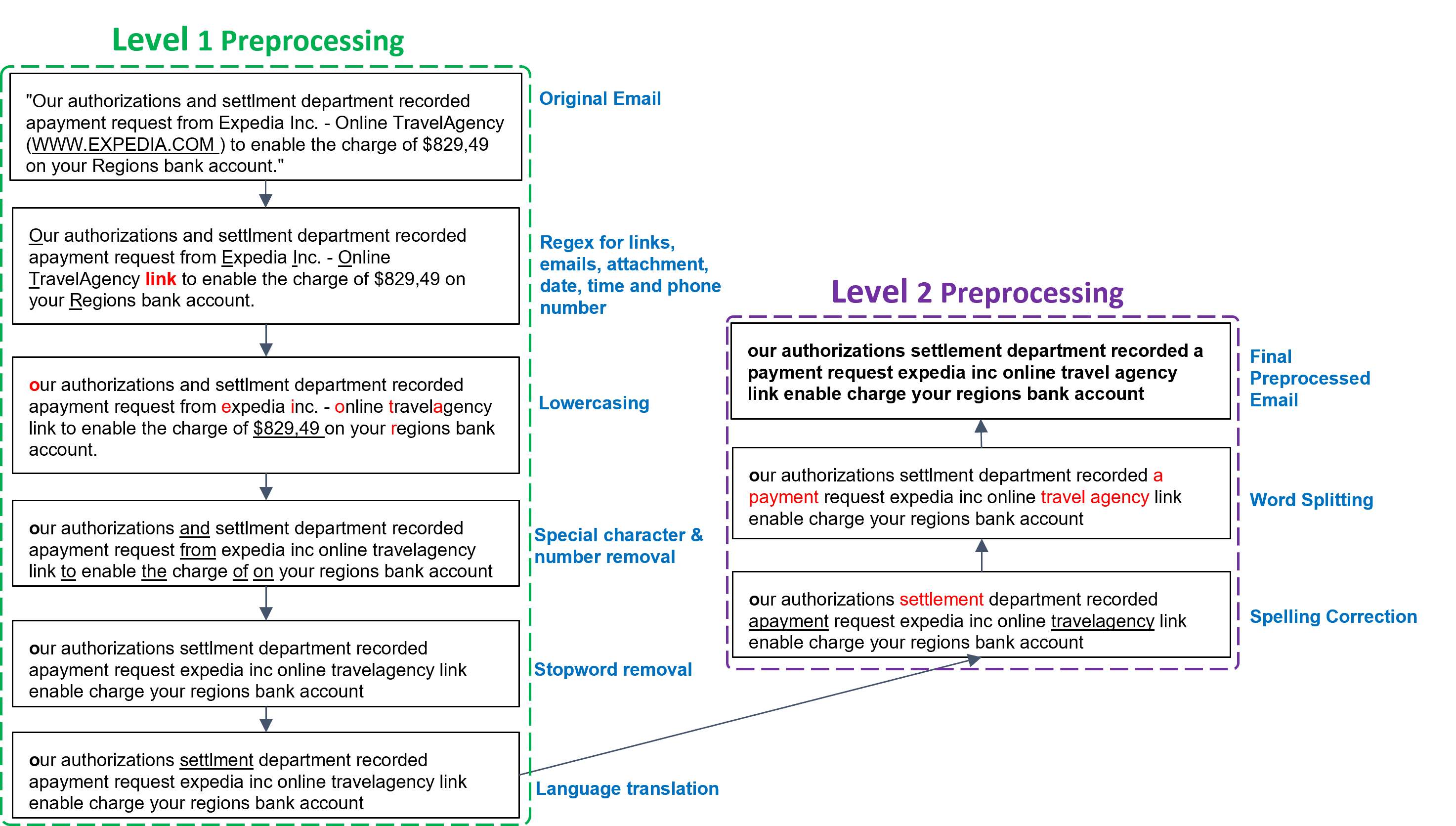}
\caption{Level 1 and 2 text preprocessing of emails}
\label{fig:email_preprocessing}
\end{figure*}

\textbf{Dataset Preprocessing}: Before feeding the emails into ML algorithms for model development, a two-level preprocessing was performed. \\
\underline{Initial Preprocessing (Level 1)}: The initial level of preprocessing focuses on cleaning and normalizing emails to ensure uniformity across all email samples. The key operations in Level 1 include lower-casing, removal of special characters, stop words, and numbers, regex substitutions of URLS, emails, and phone numbers, detection of non-english emails and their translation into English using \texttt{langdetect} \cite{langdetect} and \texttt{googletrans} \cite{googletrans} libraries, respectively.\\
\underline{Advanced Preprocessing (Level 2)}: Level 2 preprocessing targets textual anomalies specific to phishing emails, including misspellings and words concatenation. Several libraries were analyzed including \texttt{pyspellchecker} \cite{pyspellchecker}, \texttt{autocorrect} \cite{autocorrect}, \texttt{textblob} \cite{textblob}, and \texttt{symspellpy} \cite{symspellpy} for correcting spelling errors. \texttt{Textblob} was eliminated due to its limited vocabulary and limited handling of compound words. \texttt{Symspellpy} was excluded as it uses only deletion operations that restricts flexibility. \texttt{Pyspellchecker} was found to introduce unnecessary edits due to over-reliance on frequency. Finally, \texttt{Autocorrect} was selected for its edit-distance control and consistent performance. A maximum edit distance of 2 was configured for the correction engine. To resolve concatenated terms, multiple libraries were evaluated. \texttt{Wordninja} \cite{wordninja} was preferred over \texttt{wordsegment} \cite{wordsegment}, \texttt{Spacy} \cite{spacy}, and \texttt{textblob} \cite{textblob} due to its statistical segmentation through greedy algorithms based on large corpora. While \texttt{Wordsegment} offered high accuracy, \texttt{Wordninja} provided an optimal balance between speed and accuracy for large-scale preprocessing. The spelling correction and word splitting operations reduce noise and help models focus on semantic meaning.  
Figure~\ref{fig:email_preprocessing} shows the steps involved in Level 1 and 2 preprocessing of emails. 

We applied two distinct preprocessing transformations to the emails in both the training and test sets during the model development stage. The first transformation involved only Level 1 preprocessing and is referred to as the \textbf{\textit{Baseline}} dataset. The second transformation incorporated both Level 1 and 2 preprocessing steps, forming the \textbf{\textit{Advanced}} dataset, which represents our proposed approach. The same preprocessing procedures were applied to the emails in the model deployment stage. Additionally, a subset of the transformed deployment-stage emails was used to evaluate the adversarial robustness of the ML models.  
\subsection{Feature Extraction}
We employed following NLP feature extraction techniques to convert emails into vector representations while preserving the underlying semantics and patterns in textual data.
\subsubsection{Term Frequency-Inverse Document Frequency (TF-IDF)}
TF-IDF is widely used for quantifying the importance of a word in a document relative to a corpus~\cite{tfidf}. It is a product of Term-Frequency (TF) and Inverse Document Frequency (IDF).
\begin{itemize}
    \item \textbf{Term Frequency (TF)} measures the frequency of a word $w$ in an email $e$, often scaled logarithmically (Eq. \ref{eq:tf}).
    \begin{equation}
        \text{TF}(w, e) = \log(1 + \text{count}(w \in e))
        \label{eq:tf}
    \end{equation}
    \item \textbf{Inverse Document Frequency (IDF)} captures the rarity of the word $w$ across the entire corpus of emails $E$ (Eq. \ref{eq:idf}).
  \begin{equation}
       \text{IDF}(w, E) = \log\left(\frac{ \lvert E \rvert}{\lvert \{e \in E\ \text{and}\  w \in e\}\rvert}\right)
       \label{eq:idf}
    \end{equation}
    \item \textbf{TF-IDF}: The TF-IDF score for a word $w$ in an email $e$ in the email corpus $E$ is computed as follows in Eq. \ref{eq:tf-idf}:
\begin{equation}
    \text{TF-IDF}(w, e, E) = \text{TF}(w, e) \times \text{IDF}(w, E)
    \label{eq:tf-idf}
\end{equation}
\end{itemize}

The higher the TF-IDF score, the more important a word is to a specific email and less frequent in the corpus. Each email was converted into a TF-IDF feature vector. We included unigrams, bigrams, and trigrams to capture contextual information. The feature vectorizer was limited to 100 most significant terms to reduce dimensionality and overfitting. After fitting the training data, the vectorizer learned the vocabulary and IDF scores. Test data was transformed using the same vectorizer to ensure consistency between training and test dataset representations.

\subsubsection{Word2Vec}
Word2Vec is a neural embedding algorithm that maps words to dense vector representations based on contextual similarity. Unlike TF-IDF, which is frequency-based, Word2Vec leverages local context in a text to learn word embeddings~\cite{word2vec}. In our study, we trained a Word2Vec model from an open source Python library called Gensim \cite{gensim} with a vector size of 100, window size of 5 and a minimum word frequency threshold of 1. For each email, the embeddings for each word $\mathbf{v}_w$ (Eq. \ref{eq:w2vw}) in the email were averaged to yield a fixed-size vector, i.e., $\mathbf{v}_e$ (Eq. \ref{eq:w2ve}). If no known word vectors were present, a zero vector was used. Test dataset was processed using the same trained model ignoring unseen tokens.
\begin{equation}
    \mathbf{v}_w = [v_1, v_2, ..., v_{100}]
    \label{eq:w2vw}
\end{equation}

\begin{equation}
    \mathbf{v}_e = \frac{1}{n} \sum_{i=1}^{n} \mathbf{v}_{w_i}
    \label{eq:w2ve}
\end{equation}

\subsubsection{Global Vectors for Word Representation (GloVe)}
GloVe is an unsupervised learning algorithm that derives word embeddings from global co-occurrence statistics of words in a corpus~\cite{glove}. GloVe constructs a word-word co-occurrence matrix and learns embeddings by minimizing a weighted least squares objective function. Given two words $w_1$ and $w_2$ with embeddings $\mathbf{v}_1$ and $\mathbf{v}_2$, the model aims to satisfy Eq. \ref{eq:glove}:
\begin{equation}
    \mathbf{v}_1 \cdot \mathbf{v}_2 \approx \log(P(w_1, w_2))
    \label{eq:glove}
\end{equation}
where $P(w_1,\ w_2)$ is the probability of co-occurrence. We utilized pre-trained \texttt{glove.6B.100d} \cite{glovefile} embeddings, consisting of 100-dimensional vectors trained on a corpus of 6 billion words. Both training and test datasets were tokenized, and for each token found in the GloVe vocabulary, its vector representation was retrieved. Each email was represented by averaging the GloVe vectors of its constituent words. Unseen or out-of-vocabulary words were ignored, and in their absence, a zero vector was assigned.

\subsection{ML Algorithms}
After feature extraction, resulting vector representations were utilized for both training and testing for the following ML algorithms discussed as follows: 
\subsubsection{Logistic Regression (LR)}
LR is a fundamental discriminative model in supervised learning, primarily used for binary classification tasks.  
During training, the algorithm learns an optimal weight vector $\mathbf{w} = [w_1, w_2, ..., w_n]$ and a bias term $b$. Each weight $w_i$ signifies the importance of feature $x_i$ for classification between two classes, i.e., legitimate and phishing emails. The decision function computes a weighted sum $z$ of the input features using Eq. \ref{eq:lrsum}:
\begin{equation}
    z = \sum_{i=1}^{n} w_i x_i + b = \mathbf{w} \cdot \mathbf{x} + b
    \label{eq:lrsum}
\end{equation}

This value $z$ is passed through the sigmoid activation function, shown in Eq. \ref{eq:sigmoid}, to yield a class probability:
\begin{equation}
    \sigma(z) = \frac{1}{1 + \exp(-z)}
    \label{eq:sigmoid}
\end{equation}

A threshold probability of $0.5$ is typically used to predict the target class $\hat y$ (Eq. \ref{eq:lrlabel}):
\begin{equation}
    \hat y = 
    \begin{cases}
        1, & \text{if } \sigma(z) \geq 0.5 \\
        0, & \text{otherwise}
    \end{cases}
    \label{eq:lrlabel}
\end{equation}

\subsubsection{Random Forest (RF)}
RF is an ensemble learning technique that enhances prediction accuracy and robustness by constructing multiple decision trees during training. In RF with $d$ trees $\{T_1, T_2, ..., T_d\}$, each tree is trained on a random subset of data and features. For a given input, each tree $i$ outputs a class prediction $\hat y_i$ for a given set of input features. The final output, $\hat y$ is determined by majority voting using Eq. \ref{eq:rf}:

\begin{equation}
    \hat y = 
    \begin{cases}
        1, & \text{if } \sum_{i=1}^{d} 1(\hat y_i = 1) > \sum_{i=1}^{d} 0(\hat y_i = 0) \\
        0, & \text{otherwise}
    \end{cases}
    \label{eq:rf}
\end{equation}

\subsubsection{Support Vector Machine (SVM)}
SVM is a supervised learning algorithm that identifies the optimal hyperplane separating data points from different classes by maximizing the margin between the closest data points (support vectors). For a binary classification problem with input vector $\mathbf{x}$, weight vector $\mathbf{w}$, bias $b$, and label $y \in \{-1, +1\}$,  the decision function is defined as follows in Eq. \ref{eq:svm1}:
\begin{equation}
    f(\mathbf{x}) = \mathbf{w}^\top \mathbf{x} + b
    \label{eq:svm1}
\end{equation}

The classifier predicts a label $\hat y$ for a given input $\mathbf{x}$ based on the sign of the decision function (Eq. \ref{eq:svm2}):
\begin{equation}
    \hat y = 
    \begin{cases}
         1, & \text{if } f(\mathbf{x})\geq 0 \\
        -1, & \text{if } f(\mathbf{x}) < 0
    \end{cases}
    \label{eq:svm2}
\end{equation}

For non-linearly separable data, kernel functions project the input space to a higher-dimensional feature space where a linear separator is not feasible.

\subsubsection{Multi-Layer Perceptron (MLP)} MLP is a class of feed-forward artificial neural network composed of an input layer, one or more hidden layers, and an output layer. It models complex non-linear relationships between inputs and outputs. Each neuron performs a weighted sum of its inputs using Eq. \ref{eq:lrsum}. This sum is passed through a non-linear activation function (e.g., the sigmoid function in Eq. \ref{eq:sigmoid}). The model is trained using backpropagation, minimizing the loss function (e.g., binary cross-entropy in Eq. \ref{eq:cross_entropy}, which is commonly used for classification), where $N,\ y_i,\ \hat y_i$ represent number of inputs, target and predicted classes for input $i$. Each weight $w_i$ and bias $b$ are updated using gradient descent shown in Eq. \ref{eq:wbupdate}.
\begin{equation}
    \mathcal{L} = -\frac{1}{N} \sum_{i=1}^{N} [y_i \log(\hat{y}_i) + (1 - y_i) \log(1 - \hat{y}_i)]
    \label{eq:cross_entropy}
\end{equation}

\begin{equation}
\begin{aligned}
    w_i = w_i - \eta \frac{\partial \mathcal{L}}{\partial w_i}\\
    b = b - \eta \frac{\partial \mathcal{L}}{\partial b}
\end{aligned}
\label{eq:wbupdate}
\end{equation}

\subsubsection{K-Nearest Neighbors (KNN)}
KNN is an instance based learning algorithm that classifies an input sample based on the majority class among its $k$ closest training samples. It relies on distance metrics such as Euclidean, Manhattan, or Minkowski distance for neighbor computation. Given an instance $\mathbf{x}$ and training set $\mathbf{X}$, the distance of $\mathbf{x}$ from each record $i$ in $\mathbf{X}$ is calculated using Eq. \ref{eq:knn1}, where $d$ is the number of features. After identifying the $k$ nearest neighbors, the predicted class $\hat y$ is assigned based on majority voting (\ref{eq:knn2}). The choice of $k$ is crucial and small $k$ may lead to overfitting, while large $k$ may cause underfitting. 
\begin{equation}
    d(\mathbf{x}, \mathbf{X}_i) = \sqrt{\sum_{j=1}^{d} (x_j - X_{ij})^2}
    \label{eq:knn1}
\end{equation}

\begin{equation}
    \hat y = \arg \max_{c \in \{0,1\}} \sum_{n=1}^{k} 1(y_n = c)
    \label{eq:knn2}
\end{equation}

The models for these algorithms were implemented using the Scikit-learn library~\cite{scikit-learn}. It is an open-source ML library in Python that offers simple and efficient tools for data mining, data analysis, and ML workflows. Model training employed 10-fold cross-validation using the \verb|StratifiedKFold| strategy to maintain class distribution across folds. Hyperparameter tuning was conducted using \verb|GridSearchCV| to identify the optimal configurations. The best-performing models were then retained for deployment. 
\subsection{Performance Metrics for Classification Models}
To evaluate the effectiveness of the classification models, we utilized standard performance metrics commonly employed in binary classification: Accuracy (Eq. \ref{eq:accuracy}), Precision (Eq. \ref{eq:precision}), Recall (Eq. \ref{eq:recall}), F1-score (Eq. \ref{eq:f1-score}), and Receiver Operating Characteristic (ROC) curve. In our phishing email classification context, True Positives (TP) and True Negatives (TN) refer to correctly classified phishing and legitimate emails, respectively. False Positives (FP) are legitimate emails incorrectly identified as phishing, while False Negatives (FN) are phishing emails misclassified as legitimate emails.

\textbf{Accuracy}: Accuracy quantifies the overall correctness of the model by measuring the ratio of correct predictions to the total number of predictions.
\begin{equation}
\text{Accuracy} = \frac{TP + TN}{TP + FP + TN + FN}
\label{eq:accuracy}
\end{equation}

\textbf{Precision}: Precision indicates the proportion of true phishing emails among all emails predicted as phishing. It is particularly useful when the cost of false positives is high.
\begin{equation}
\text{Precision} = \frac{TP}{TP + FP}
\label{eq:precision}
\end{equation}

\textbf{Recall}: Recall (also known as Sensitivity or True Positive Rate) measures the ability of the model to correctly identify actual phishing emails.
\begin{equation}
\text{Recall} = \frac{TP}{TP + FN}
\label{eq:recall}
\end{equation}

\textbf{F1-score}: F1-score is the harmonic mean of Precision and Recall, providing a single measure that balances precision and recall. It serves as a key performance indicator for imbalanced datasets, such as in our case, where accuracy alone is insufficient to evaluate the model's performance.
\begin{equation}
\text{F1-score} = 2 \times \frac{Precision \times Recall}{Precision + Recall}
\label{eq:f1-score}
\end{equation}

\textbf{Receiver Operating Characteristic (ROC) Curve}: The  ROC curve is a graphical tool used to assess the performance of a binary classifier across different threshold values. It plots the True Positive Rate (TPR) against the False Positive Rate (FPR) as defined in Eq. \ref{eq:tpr_fpr}.
\begin{equation}
\text{TPR} = \frac{TP}{TP + FN}, \quad \text{FPR} = \frac{FP}{FP + TN}
\label{eq:tpr_fpr}
\end{equation}
 
The Area Under the ROC Curve (AUC-ROC) is often used as a scalar measure to summarize the model's performance.

\subsection{Model Development and Deployment Results}
The performance is compared for the models trained with Baseline dataset that uses Level 1 preprocessing and Advanced dataset with both Level 1 and 2 preprocessing. The results are split between model development and model deployment settings.
\subsubsection{Model Performance on Baseline Dataset}
\begin{figure*}[htbp]
    \centering
    \begin{subfigure}[b]{0.32\linewidth}
        \centering
        \includegraphics[width=\linewidth]{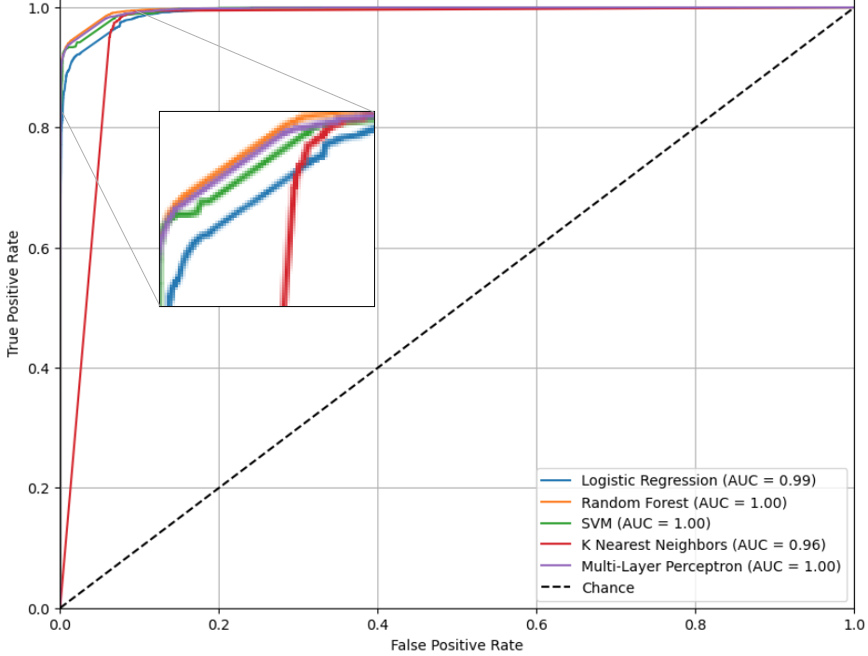}
        \caption{TF-IDF with Level 1}
        \label{fig:roc_tf_idf_1}
    \end{subfigure}
    \hfill
    \begin{subfigure}[b]{0.32\linewidth}
        \centering
        \includegraphics[width=\linewidth]{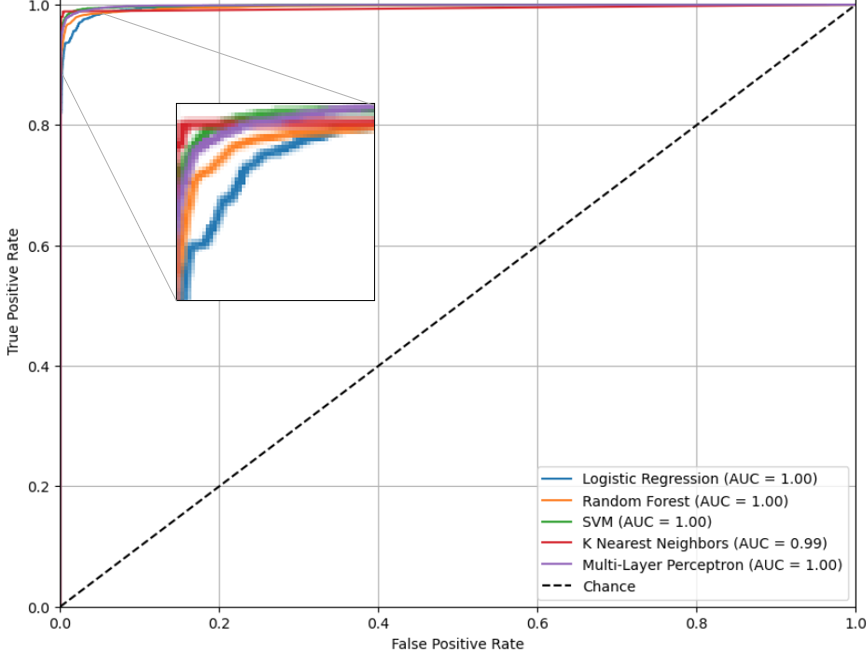}
        \caption{Word2Vec with Level 1}
        \label{fig:roc_word2vec_1}
    \end{subfigure}
    \hfill
    \begin{subfigure}[b]{0.32\linewidth}
        \centering
        \includegraphics[width=\linewidth]{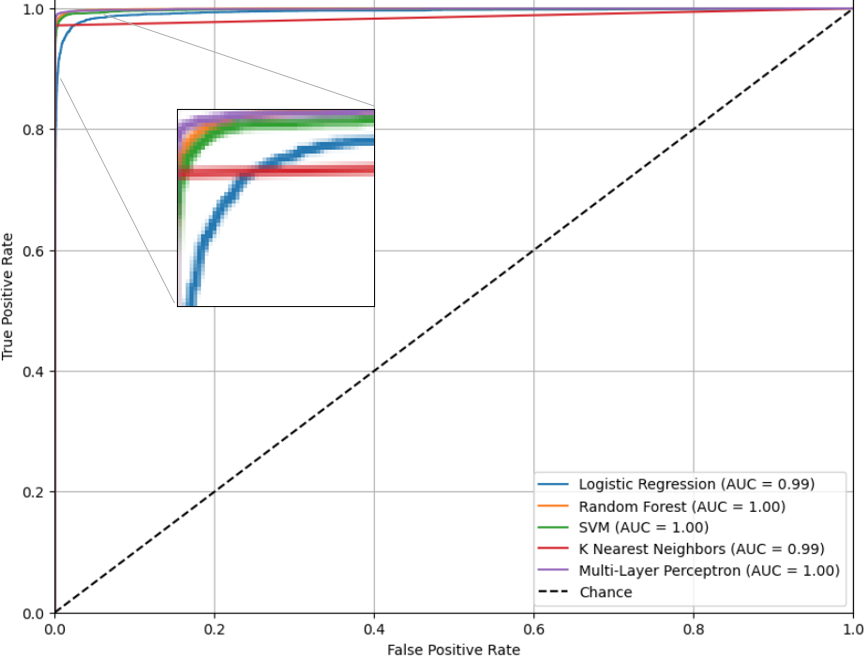}
        \caption{Glove with Level 1}
        \label{fig:roc_glove_1}
    \end{subfigure}
    \vspace{1em} 
    \begin{subfigure}[b]{0.32\linewidth}
        \centering
        \includegraphics[width=\linewidth]{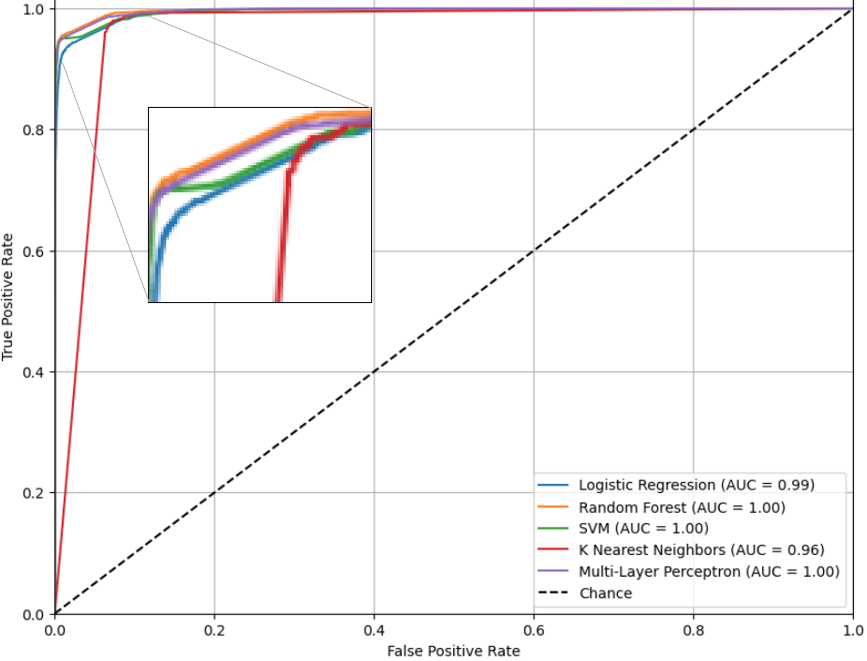}
        \caption{TF-IDF with Level 1 \& 2}
        \label{fig:roc_tf_idf_2}
    \end{subfigure}
    \hfill
    \begin{subfigure}[b]{0.32\linewidth}
        \centering
        \includegraphics[width=\linewidth]{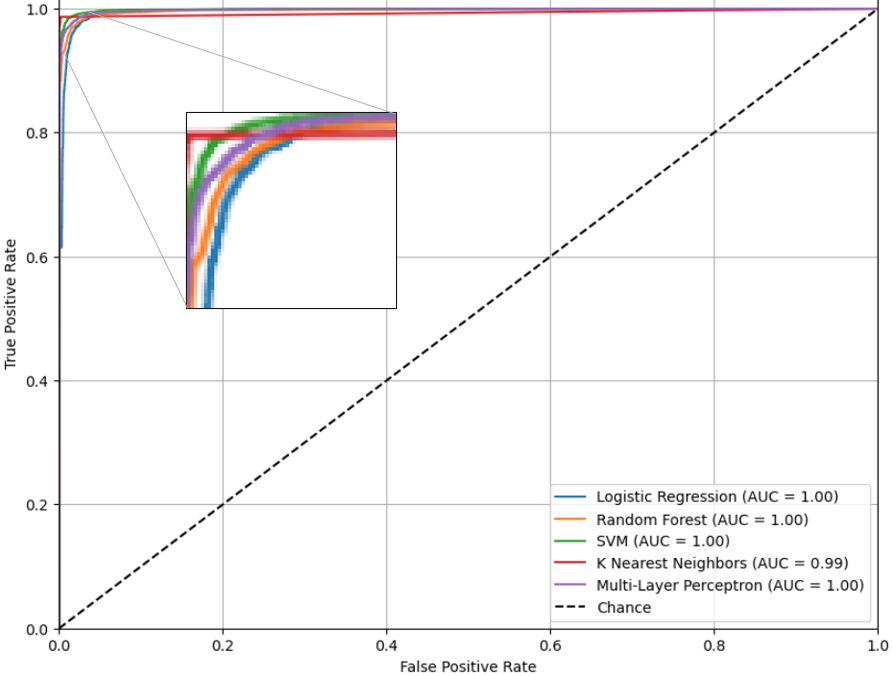}
        \caption{Word2Vec with Level 1 \& 2}
        \label{fig:roc_word2vec_2}
    \end{subfigure}
    \hfill
    \begin{subfigure}[b]{0.32\linewidth}
        \centering
        \includegraphics[width=\linewidth]{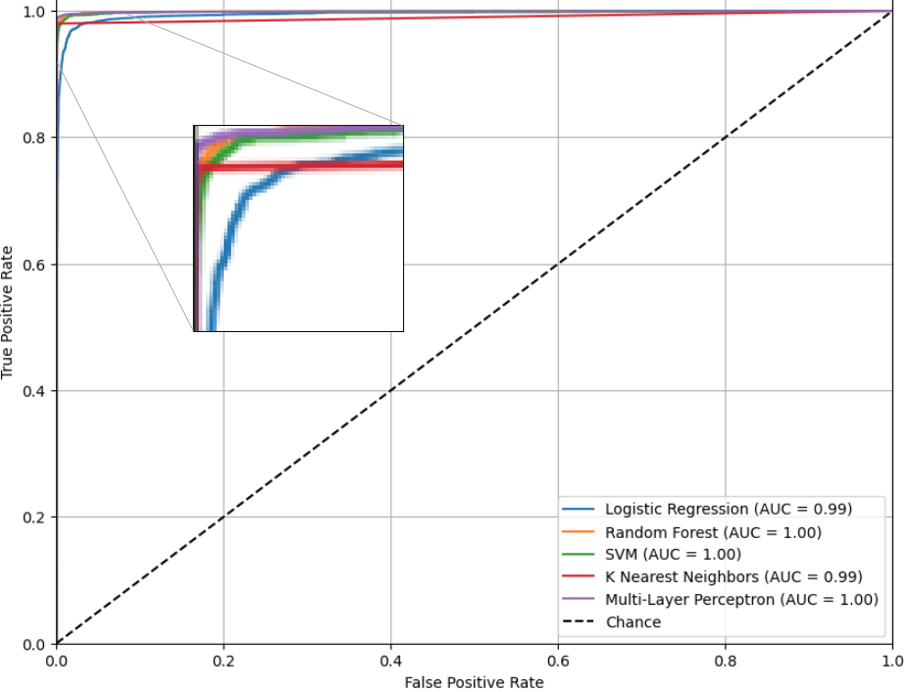}
        \caption{Glove with Level 1 \& 2}
        \label{fig:roc_glove_2}
    \end{subfigure}

    \caption{ROC curves for models trained with Baseline dataset (Level 1 preprocessing) and Advanced dataset (Level 1 and 2 preprocessing)}
    \label{fig:roc_curves}
\end{figure*}

In the development phase, among models using TF-IDF, RF achieved highest F1-score of 95.19\%, followed closely by MLP (95.02\%) and SVM (94.97\%). The least performance with TF-IDF was observed for KNN with an F1-score of 85.42\%. For Word2Vec embeddings, MLP achieved the best performance with an F1-score of 99.19\%, while LR scored the lowest with 97.53\%. With GloVe embeddings, MLP achieved highest F1-score of 98.65\%, and LR achieved the lowest (95.02\%). Overall, MLP was the best-performing model across all feature types for Baseline dataset, while KNN with TF-IDF performed the least. Figure \ref{fig:roc_tf_idf_1}-\ref{fig:roc_glove_1} show ROC curves and Table ~\ref{tab:dev_results_dataset1} summarizes the results for all the ML models developed. 

Results showed more variation in model deployment setting for Baseline dataset. With TF-IDF, SVM  with an F1-score of 79.90\% outperformed other models marginally, while KNN again showed the least performance (69.62\%). With Word2Vec, MLP had the best performance with an F1-score of 86.64\%, while KNN and RF have least F-scores of 81.44\% and 80.57\%, respectively. For GloVe, the SVM model performed the best with an F1-score of 81.10\% and KNN the least with an F1-score of 64.35\%. Table~\ref{tab:deploy_results_dataset1} shows the model deployment results.

\begin{table}[h!]
\centering
\caption{Development results for models trained with baseline Dataset}
\label{tab:dev_results_dataset1}
\begin{tabular}{|c|c|c|c|c|c|c|}
\hline
\textbf{Feature Type} & \textbf{Model} & \textbf{Train Acc.} & \textbf{Test Acc.} & \textbf{Precision} & \textbf{Recall} & \textbf{F1-score} \\
\hline

\multirow{5}{*}{TF-IDF}
& LR  & 97.12\% & 97.26\% & 96.09\% & 88.92\% & 92.37\% \\
& SVM & 98.36\% & 98.18\% & 98.21\% & 91.94\% & 94.97\% \\
& RF  & 98.45\% & 98.26\% & 98.35\% & 92.23\% & \textbf{95.19}\% \\
& KNN & 94.38\% & 93.78\% & 75.90\% & 97.69\% & 85.42\% \\
& MLP & 98.41\% & 98.20\% & 98.02\% & 92.20\% & 95.02\% \\
\hline

\multirow{5}{*}{Word2Vec}
& LR  & 99.20\% & 99.08\% & 98.02\% & 97.05\% & 97.53\% \\
& SVM & 99.91\% & 99.67\% & 99.68\% & 98.54\% & 99.11\% \\
& RF  & 99.99\% & 99.32\% & 99.58\% & 96.78\% & 98.16\% \\
& KNN & 99.99\% & 99.52\% & 99.94\% & 97.49\% & 98.70\% \\
& MLP & 99.97\% & 99.70\% & 99.63\% & 98.76\% & \textbf{99.19}\% \\
\hline

\multirow{5}{*}{GloVe}
& LR  & 98.08\% & 98.16\% & 96.09\% & 93.96\% & 95.02\% \\
& SVM & 99.19\% & 99.20\% & 98.41\% & 97.29\% & 97.85\% \\
& RF  & 99.98\% & 99.17\% & 99.34\% & 96.23\% & 97.76\% \\
& KNN & 99.98\% & 99.08\% & 99.90\% & 95.20\% & 97.49\% \\
& MLP & 99.89\% & 99.50\% & 99.64\% & 97.67\% & \textbf{98.65}\% \\
\hline
\end{tabular}
\end{table}

\begin{table}[h!]
\centering
\caption{Deployment results for models trained with baseline dataset}
\label{tab:deploy_results_dataset1}
\footnotesize
\begin{tabular}{|c|c|c|c|c|c|}
\hline
\textbf{Model}  & \textbf{Accuracy} & \textbf{Precision} & \textbf{Recall} & \textbf{F1-score} \\
\hline
\multicolumn{5}{|c|}{TF-IDF} \\
\hline
LR  & 93.78\% & 90.41\% & 70.11\% & 78.98\% \\
SVM  & 93.98\% & 90.22\% & 71.69\% & \textbf{79.90}\% \\
RF  & 93.34\% & 89.35\% & 68.20\% & 77.35\% \\
KNN  & 88.22\% & 61.08\% & 80.93\% & 69.62\% \\
MLP  & 93.14\% & 85.04\% & 71.43\% & 77.64\% \\
\hline
\multicolumn{5}{|c|}{Word2Vec} \\
\hline
LR  & 94.56\% & 81.99\% & 86.30\% & 84.09\% \\
SVM  & 93.54\% & 75.48\% & 90.75\% & 82.42\% \\
RF  & 93.09\% & 75.82\% & 85.95\% & 80.57\% \\
KNN  & 93.83\% & 81.59\% & 81.29\% & 81.44\% \\
MLP  & 95.37\% & 83.44\% & 90.09\% & \textbf{86.64}\% \\
\hline
\multicolumn{5}{|c|}{GloVe} \\
\hline
LR  & 93.66\% & 83.88\% & 76.70\% & 80.13\% \\
SVM  & 94.11\% & 87.15\% & 75.84\% & \textbf{81.10}\% \\
RF  & 92.19\% & 79.11\% & 72.21\% & 75.50\% \\
KNN  & 90.74\% & 89.92\% & 50.01\% & 64.35\% \\
MLP & 93.52\% & 87.91\% & 70.91\% & 78.50\% \\
\hline
\end{tabular}
\end{table}

\subsubsection{Model Performance on Advanced Dataset}
With TF-IDF, RF model performed the best with an F1-score of 96.10\%, followed closely by MLP (96.04\%) and SVM (95.95\%). KNN achieved the least F1-score of 85.34\%. With Word2Vec, MLP achieved the highest F1-score of 99.49\%, and LR had the lowest score of 97.67\%. For GloVe, MLP achieved the best score of 99.02\%, while LR had the lowest score of 94.88\%. Notably, the overall performance for all models increased slightly compared to their counterparts with Baseline dataset, with SVM and MLP being the top-performing models. Figure \ref{fig:roc_tf_idf_2}-\ref{fig:roc_glove_2} show ROC curves and Table ~\ref{tab:dev_results_dataset2} summarizes the results for all the ML models developed. 
\begin{table}[h!]
\centering
\caption{Development results for models trained with Advanced dataset \cite{kulal}}
\label{tab:dev_results_dataset2}
\begin{tabular}{|c|c|c|c|c|c|c|}
\hline
\textbf{Feature Type} & \textbf{Model} & \textbf{Train Acc.} & \textbf{Test Acc.} & \textbf{Precision} & \textbf{Recall} & \textbf{F1-score} \\
\hline

\multirow{5}{*}{TF-IDF}
& \cellcolor{lightgray} LR  & 97.66\% & 97.80\% & 94.56\% & 91.45\% & 93.93\% \\
& \cellcolor{lightgray} SVM & 98.78\% & 98.52\% & 97.99\% & 94.00\% & 95.95\% \\
& \cellcolor{lightgray} RF  & 98.87\% & 98.57\% & 98.25\% & 94.05\% & \textbf{96.10}\% \\
& KNN & 94.39\% & 93.75\% & 75.83\% & 97.81\% & 85.34\% \\
& \cellcolor{lightgray} MLP & 98.84\% & 98.55\% & 98.28\% & 93.90\% & 96.04\% \\ \hline

\multirow{5}{*}{Word2Vec}
& \cellcolor{lightgray} LR  & 99.04\% & 99.14\% & 98.19\% & 97.16\% & 97.67\% \\
& \cellcolor{lightgray} SVM & 99.86\% & 99.77\% & 99.74\% & 99.05\% & 99.39\% \\
& \cellcolor{lightgray} RF  & 99.99\% & 99.41\% & 99.72\% & 97.13\% & 98.41\% \\
& KNN & 99.99\% & 99.49\% & 99.91\% & 97.38\% & 98.62\% \\
& \cellcolor{lightgray} MLP & 99.99\% & 99.81\% & 99.62\% & 99.38\% & \textbf{99.49}\% \\
\hline

\multirow{5}{*}{GloVe}
& LR  & 98.02\% & 98.09\% & 95.13\% & 94.63\% & 94.88\% \\
& \cellcolor{lightgray} SVM & 99.25\% & 99.26\% & 98.40\% & 97.61\% & 98.01\% \\
& \cellcolor{lightgray} RF  & 99.99\% & 99.18\% & 99.43\% & 96.20\% & 97.79\% \\
& \cellcolor{lightgray} KNN & 99.99\% & 99.28\% & 99.73\% & 96.41\% & 98.04\% \\
& \cellcolor{lightgray} MLP & 99.89\% & 99.64\% & 99.72\% & 98.34\% & \textbf{99.02}\% \\ \hline
\end{tabular}
\end{table}

\begin{table}[h!]
\centering
\caption{Deployment results for models trained with Advanced dataset \cite{kulal}}
\label{tab:deploy_results_dataset2}
\footnotesize
\begin{tabular}{|c|c|c|c|c|c|}
\hline
\textbf{Model}  & \textbf{Accuracy} & \textbf{Precision} & \textbf{Recall} & \textbf{F1-score} \\
\hline
\multicolumn{5}{|c|}{TF-IDF} \\
\hline
\rowcolor{lightgray} LR  & 94.02\% & 90.44\% & 71.69\% & 79.99\% \\
\rowcolor{lightgray} SVM & 94.22\% & 88.76\% & 74.79\% & 81.18\% \\
\rowcolor{lightgray} RF  & 94.82\% & 91.45\% & 74.10\% & \textbf{82.66}\% \\
\rowcolor{lightgray} KNN & 89.53\% & 65.38\% & 79.00\% & 71.55\% \\
\rowcolor{lightgray} MLP & 94.36\% & 89.22\% & 75.26\% & 81.65\% \\
\hline
\multicolumn{5}{|c|}{Word2Vec} \\
\hline
LR  & 94.07\% & 79.82\% & 86.24\% & 82.91\% \\
\rowcolor{lightgray} SVM & 94.16\% & 77.10\% & 92.45\% & 84.08\% \\
RF  & 93.83\% & 84.90\% & 76.64\% & 80.56\% \\
KNN & 93.49\% & 92.62\% & 66.24\% & 77.24\% \\
MLP & 94.26\% & 77.16\% & 93.12\% & \textbf{84.39}\% \\
\hline
\multicolumn{5}{|c|}{GloVe} \\
\hline
LR  & 92.20\% & 75.09\% & 79.58\% & 77.27\% \\
\rowcolor{lightgray} SVM & 93.65\% & 78.48\% & 85.34\% & 81.77\% \\
RF  & 91.12\% & 70.75\% & 79.67\% & 74.94\% \\
\rowcolor{lightgray} KNN & 91.55\% & 77.88\% & 68.87\% & 73.10\% \\
\rowcolor{lightgray} MLP & 93.79\% & 79.10\% & 85.32\% & \textbf{82.09}\% \\
\hline
\end{tabular}
\end{table}

In model deployment setting, results confirmed the effectiveness of Level 1 and 2 preprocessing used in Advanced dataset. For TF-IDF, RF performed best with an F1-score of 82.66\%, and KNN remained the lowest with a score of 71.55\%. Word2Vec showed strong performance for MLP (84.39\%) and SVM (84.08\%), while KNN showed weak behavior with a F1-score of 77.24\%. GloVe embeddings continued the trend of stable performance, where MLP had the highest F1-score of 82.09\%, and KNN achieved the lowest score of 73.10\%. Table~\ref{tab:deploy_results_dataset2} shows the model deployment results.

\subsubsection{Summary of Model Development and Deployment Results}
We observed that models trained on Advanced dataset—which includes additional preprocessing steps such as spelling correction and word splitting—consistently outperformed their Baseline dataset counterparts in both development and deployment evaluations. The highlighted rows in Table~\ref{tab:dev_results_dataset2} and Table~\ref{tab:deploy_results_dataset2} indicate the models that achieved superior performance when trained with Advanced dataset. These results validate our \textbf{RQ1}, demonstrating that ML-based phishing detection models benefit from enhanced preprocessing techniques, such as spelling correction and word splitting, in effectively identifying phishing emails.

\subsubsection{Computational Time and Hyperparameters}
To train the ML models, a workstation with AMD Ryzen 9 5900X 12-Core Processor, 3.7 GHz, 12 Cores(s), 24 Logical Processors, and 32 GB RAM was used. For Baseline dataset, the highest training time was recorded for the SVM model with TF-IDF at 41900.12 seconds, due to the high dimensionality of the vectors. The fastest training was achieved by LR with Glove, requiring only 268.66 seconds. For testing, SVM with TF-IDF had the slowest time (19.62 seconds), while the fastest was MLP with GloVe (2.17 seconds). For Advanced dataset, the longest training time was observed with the SVM model using TF-IDF, taking 36160.01 seconds, whereas the shortest training time was LR with Glove at 292.69 seconds. For testing, SVM with TF-IDF embeddings was the slowest at 17.24 seconds, while LR with GloVe was the most efficient, completing prediction in 2.16 seconds. The hyperparameters for each ML model were carefully tuned using Grid Search for both Baseline and Advanced datasets. As observed in Tables~\ref{tab:dev_results_dataset1}–\ref{tab:deploy_results_dataset2}, the ML models trained using Word2Vec embeddings consistently outperformed those using TF-IDF and GloVe features in both development and deployment settings. Consequently, Table~\ref{tab:hyperparameters} presents the hyperparameters used for the models trained with Word2Vec embeddings only for the sake of conciseness.

\begin{table}[!h]
\centering
\caption{Hyperparameters for models trained with Word2Vec embedding}
\label{tab:hyperparameters}
\scriptsize
\begin{tabular}{|c|c|p{5cm}|}
\hline
\textbf{Dataset} & \textbf{Model} & \textbf{Hyperparameters} \\ \hline

\multirow{5}{*}{Baseline Dataset}
& LR & C=10, penalty=l2, max\_iter=200, solver=saga \\ \cline{2-3}
& RF & max\_features=sqrt, min\_samples\_leaf=1, min\_samples\_split=2 \\ \cline{2-3}
& MLP & activation=relu, alpha=0.0001, hidden\_layer\_sizes=(100,), learning\_rate=constant, solver=adam \\ \cline{2-3}
& SVM & C=1.0, gamma=scale, kernel=rbf \\ \cline{2-3}
& KNN & metric=euclidean, n\_neighbors=3, weights=distance \\ \hline

\multirow{5}{*}{Advanced Dataset}
& LR & C=100, penalty=l2, max\_iter=100, solver=lbfgs \\ \cline{2-3}
& RF & max\_features=sqrt, min\_samples\_leaf=1, min\_samples\_split=2 \\ \cline{2-3}
& MLP & activation=relu, alpha=0.0001, hidden\_layer\_sizes=(100,), learning\_rate=adaptive, solver=adam \\ \cline{2-3}
& SVM & C=10, gamma=scale, kernel=rbf \\ \cline{2-3}
& KNN & metric=euclidean, n\_neighbors=3, weights=distance \\ \hline

\end{tabular}
\end{table}
\section{Adversarial Robustness of Phishing Detection Models}
\label{sec:adversarial_analysis}
Understanding and detecting adversarial behavior are critical in the evolving landscape of phishing attacks. As phishing tactics rapidly advance, traditional detection methods increasingly fall short. In this work, we assess the adversarial robustness of our phishing detection model by generating and analyzing adversarial attacks that introduce subtle perturbations to the input data. These attacks are designed to degrade model performance and expose potential vulnerabilities. Phishing detectors are susceptible to adversarial attacks targeting various components, particularly text-based content. Subtle manipulations to subject lines, email bodies, and embedded URLs can exploit these systems. URL-based attacks leverage techniques such as typo-squatting, deceptive subdomains, Internationalized Domain Names (IDNs), and URL encoding to evade detection. Additionally, email header spoofing manipulates sender identities by exploiting weaknesses in SPF (Sender Policy Framework) and DKIM (DomainKeys Identified Mail) authentication. 

Our work focuses primarily on character-level text attacks and text-based content spoofing, which maintain human readability while undermining model performance. To preserve realism and semantic integrity, the adversarial modifications are constrained by factors such as Levenshtein edit distance and stop-word protection. Model robustness is quantified by comparing classification performance on adversarially perturbed inputs against baseline performance on clean data. Furthermore, we evaluate the effectiveness of our custom preprocessing in mitigating adversarial impacts. 
\subsection{TextAttack Framework for Adversarial Attacks}
We employ the \texttt{TextAttack} \cite{textattack} Python framework, which provides an extensible API for adversarial attacks, adversarial training, and data augmentation in natural NLP. It supports key components such as datasets, models, loggers, transformations, constraints, search methods, and goal functions enabling systematic construction of attack pipelines. The following components are fundamental in building adversarial attacks in \texttt{TextAttack}: 

\textit{\ul{Goal Function}} defines the attack objective and determines attack success. Two commonly used goal functions are: i) \textbf{Targeted Classification} that forces the model to misclassify the input into a specific target label. ii) \textbf{Untargeted Classification} that forces the model to misclassify the input into a label other than the original. 

\textit{\ul{Constraints}} restrict the nature and extent of perturbations to preserve semantic integrity. They include: i) \textbf{StopwordModification} prevents the modification of stop words during adversarial generation. ii) \textbf{RepeatModification} ensures that no word is modified more than once. 

\textit{\ul{Transformations}} define operations used to manipulate the text. The following character-level transformations are utilized: i) \textbf{Character Insertion} inserts random or special characters within a word. ii) \textbf{Character Deletion} deletes characters to simulate typographical errors. iii) \textbf{Neighboring Character Swap} swaps adjacent characters to mimic common human typos. iv) \textbf{Character Substitution} replaces characters with others to introduce noise. v) \textbf{Homoglyph Word Swap} substitutes characters with visually similar alternatives. vi) \textbf{Keyboard-Based Character Swap} uses keyboard adjacency to introduce likely typing mistakes. 

\textit{\ul{Search Methods}} guide the exploration of transformation space under constraints to achieve attack goals. Two methods employed include: i) \textbf{GreedySwap} performs a greedy evaluation of word substitutions to maximize degradation in model prediction confidence. ii) \textbf{GreedySwapWIR (Word Importance Ranking)} prioritizes modifications to words with the greatest influence on the model's output.\\
\textit{\ul{Model Wrapper}}: \texttt{TextAttack} supports both pre-trained and custom models. To attack a local model, a wrapper must be created where the model exposes a \texttt{\_\_call\_\_()} method for inference and a tokenizer with an \texttt{encode()} method to convert text to input tensors. The output should be a probability distribution or logits compatible with the attack's goal function. Datasets are also loaded as input-output pairs. 

\texttt{TextAttack} recipes provide ready-to-use attack configurations containing all four components including  goal function, constraints, transformations, and search strategy making it easy to construct adversarial pipelines. To rigorously test the adversarial robustness of our phishing classification model, we used three widely used attack recipes - \texttt{DeepWordBug}, \texttt{Pruthi}, and \texttt{TextBugger}, along with a custom-designed attack recipe. 

\texttt{DeepWordBug} \cite{gao2018deepwordbug} is a character-level adversarial attack recipe that supports both targeted and untargeted classification goals. It introduces perturbations constrained by a Levenshtein edit distance of at most 30, which ensures the adversarial input differs from the original by no more than 30 single-character edits. Additional constraints prevent modification of stop words and ensure that each word is altered only once. The transformations employed include character insertion, character deletion, neighboring character swap, and character substitution. The recipe uses the GreedySwapWIR search method.

\texttt{Pruthi} is an untargeted attack inspired by the work of Pruthi et al.~\cite{pruthi2019combating}, which simulates real-world typographical errors using minimal character-level changes. It applies transformations such as neighboring character swap, character insertion, deletion, and keyboard-based swaps, mimicking user typing errors. The constraints enforce a minimum word length of four characters for perturbation eligibility and limit the number of words that can be perturbed in a given input. The maximum number of character change that can be made in a word is 1. Stop words are protected from modification, and repeated alterations to the same word are not allowed. The GreedySwap search strategy is utilized.

\texttt{TextBugger}, originally proposed in~\cite{li2018textbugger}, was adapted for this study by eliminating the use of embedding-based word swaps and to focus only on character-level manipulations. The attack is targeted, aiming to bring in a specific misclassification. It enforces a semantic similarity constraint based on cosine similarity of sentence embeddings using the Universal Sentence Encoder (USE), ensuring the adversarial example remains semantically close to the original input with a threshold (typically set to 0.8). The transformations include character insertion, deletion, swap, and substitution. Like DeepWordBug, it uses the GreedySwapWIR search method.

\begin{table*}[htbp]
\centering
\caption{Comparison of adversarial attack recipes}
\label{tab:attack_recipes}
\renewcommand{\arraystretch}{1.4}
\scriptsize
\begin{tabular}{|p{2.5cm}|p{2.5cm}|p{3.5cm}|p{3.2cm}|p{2.5cm}|}
\hline
\textbf{Attack Name} & \textbf{Goal Function} & \textbf{Constraints} & \textbf{Transformations} & \textbf{Search Method} \\
\hline
DeepWordBug & Targeted / Untargeted & Levenshtein $\leq 30$, Stopword and Repeat Modification constraints & Insertion, Deletion, Swap, Substitution & Greedy-WIR \\
\hline
Pruthi & Untargeted & Min word length $\geq 4$, max words perturbed, Stopword and Repeat Modification & Insertion, Deletion, Swap, Keyboard-based Swap & Greedy \\
\hline
TextBugger & Targeted & USE cosine similarity $\geq 0.8$, Stopword and Repeat Modification & Insertion, Deletion, Swap, Substitution, Homoglyph & Greedy-WIR \\
\hline
\textbf{Custom Attack} & Untargeted & Levenshtein $\leq 16$, Stopword and Repeat Modification & Homoglyph, Insertion, Deletion, Substitution & Greedy-WIR \\
\hline
\end{tabular}
\end{table*}
The \textbf{custom attack} that we developed in this work supports untargeted classification and is designed to simulate subtle yet impactful adversarial noise. It imposes a Levenshtein edit distance constraint with a maximum threshold of 16 and restricts changes to stop words. The transformations include homoglyph word substitution, random character insertion, substitution, and deletion. This combination allows the attack to explore a wide range of perturbations while maintaining textual coherence. The search method employed is GreedySwapWIR. Table~\ref{tab:attack_recipes} lists the comparison of different attack recipes used in our work. Table \ref{tab:adversarial_samples} shows adversarial samples generated using different attack recipes. 
\begin{table}[t]
\centering
\caption{Examples of transformed email tokens by different attack recipes}
\label{tab:adversarial_samples}
\begin{tabular}{|p{2.8cm}|p{5cm}|}
\hline
\textbf{Attack Type} & \textbf{Transformed Email} \\
\hline
DeepWordBug & acc0unt l0cked react1vated replly conntact l1nk unsubscr1be  unsubscr1be\\
\hline
Pruthi & acccount loccked reactiavted prelly conntact unisubscribe unisubscribe\\
\hline
TextBugger & acount reactivated rep1y conact linnk unsubscrebe unsubscrebe\\
\hline
Custom Attack & acc0unt l0cked react1vated replly conntact l1nk unsubscr1be  unsubscr1be\\
\hline
\end{tabular}
\end{table}

\subsection{Adversarial Attack Approach}
\begin{figure*}[h]
\centering
\includegraphics[width=0.90\textwidth]{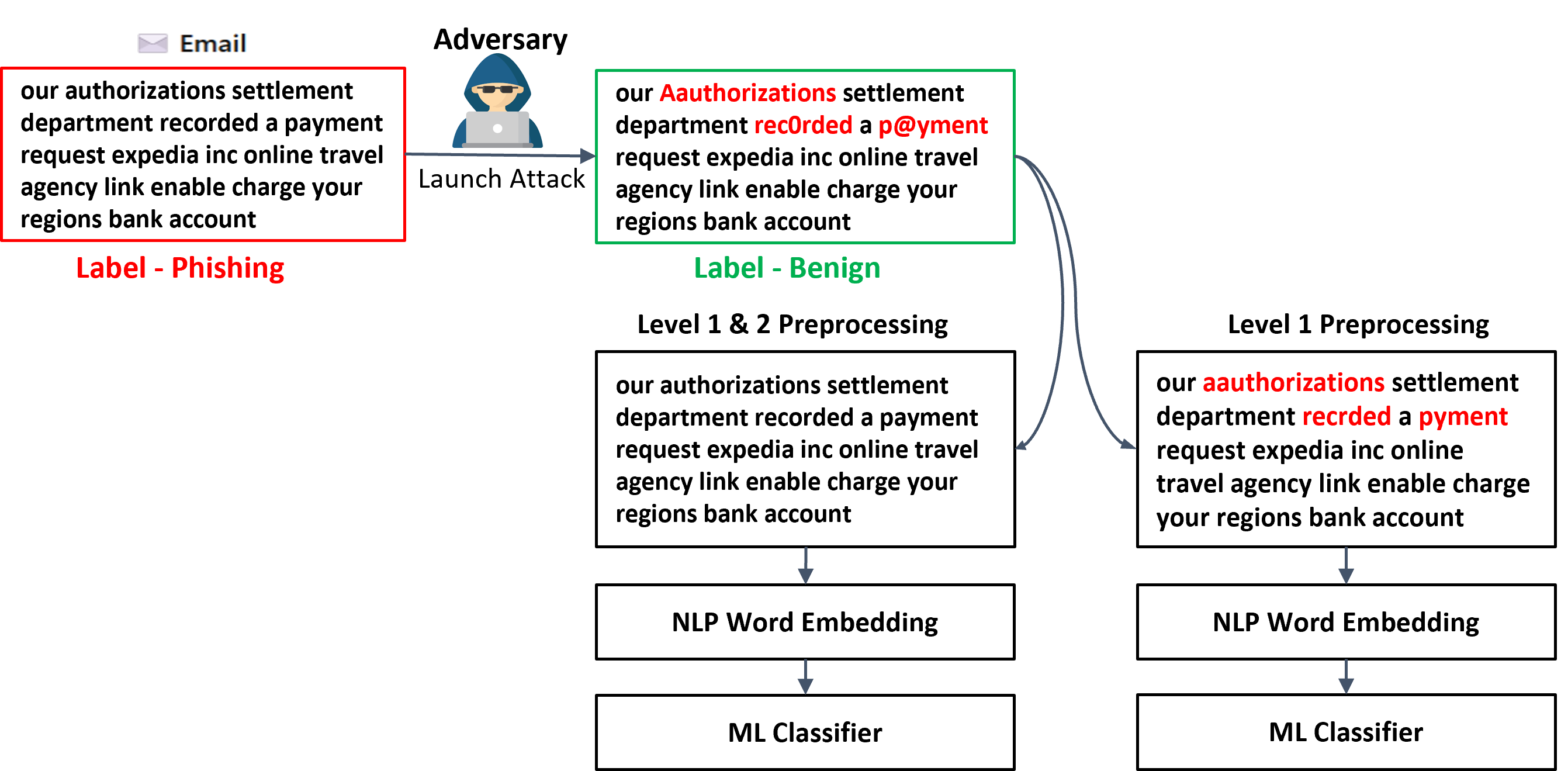}
\caption{Evaluation of adversarial robustness of phishing detection systems}
\label{fig:adversarial_approach}
\end{figure*}

To evaluate the vulnerability of the classification model under adversarial conditions, an open-box targeted adversarial attack was implemented using the \texttt{TextAttack} framework. It is targeted because it focuses solely on phishing emails and attempts to transform them in such a way that the model incorrectly predicts them into legitimate. Since \texttt{TextAttack} expects a standardized interface, a custom model wrapper is defined to ensure compatibility with the attack framework. The raw input texts are converted into token vectors using the tokenizer, after which the adversarial attack is constructed by assembling four main components: transformations, constraints, a search method, and a goal function. These components are organized into an \texttt{Attack} object. Attack arguments such as the number of samples, checkpoint saving, and logging parameters are specified using \texttt{AttackArgs}. An \texttt{Attacker} instance is created using the attack configuration, dataset, and arguments. Figure~\ref{fig:adversarial_approach} shows the approach used for evaluating the adversarial robustness of phishing detection systems for both the email preprocessing techniques. 

The attack has full access to the model’s architecture and parameters, allowing for gradient-aware transformations that maximize model misclassification. Post-attack, the perturbed adversarial samples are subjected to a preprocessing pipeline. This includes standard NLP techniques including Level 1 and both Level 1 and Level 2 techniques separately. Feature extraction techniques such as TF-IDF, Word2Vec and Glove are then reapplied to convert the cleaned adversarial texts into numerical vectors suitable for the model input. Finally, the classifier is tested on this modified dataset to observe its resilience against adversarial manipulation.  
\subsection{Metrics for Adversarial Attack Evaluation}
The following metrics are used to evaluate the performance and robustness of ML models under adversarial attacks.
\begin{itemize}
\item \textbf{Number of Successful Attacks:} The total number of inputs for which the adversarial attack successfully altered the model's prediction from phishing to legitimate. 
\item \textbf{Number of Failed Attacks:} The number of attempted attacks that failed to change the model's original prediction. 
\item \textbf{Number of Skipped Attacks:} Inputs where the attack was not attempted. This typically occurs when the model’s prediction was already incorrect on the original input or when the input did not meet the attack criteria.
\item \textbf{Original Accuracy:} The classification accuracy of the model on clean, unperturbed inputs defined in Eq. \ref{eq:orig_accuracy}. 
\begin{equation}
\small \textstyle \text{Original Accuracy (\%)} = \frac{\text{Correct Predictions on Original Inputs}}{\text{Total Inputs}} \times 100
\label{eq:orig_accuracy}
\end{equation}
\item \textbf{Accuracy Under Attack (\%):} The accuracy of the model when evaluated on adversarial examples defined in Eq. \ref{eq:accuracy_under_attack}.
\begin{equation}
\scriptsize \textstyle \text{Accuracy Under Attack (\%)} = \frac{\text{Correct Predictions on Adversarial Inputs}}{\text{Total Inputs}} \times 100
\label{eq:accuracy_under_attack}
\end{equation}
\item \textbf{Attack Success Rate (\%):} The proportion of successful attacks over all attempted attacks defined in Eq. \ref{eq:attack_success_rate}.
\begin{equation}
\small \textstyle \text{Attack Success Rate (\%)} = \frac{\text{Successful Attacks}}{\text{Successful Attacks + Failed Attacks}} \times 100
\label{eq:attack_success_rate}
\end{equation}
\item \textbf{Average Perturbed Word (\%):} The average percentage of words in the input that were modified to generate successful adversarial examples defined in Eq. \ref{eq:perturbed_word}.
\begin{equation}
\small \textstyle \text{Perturbed Word (\%)} = \frac{\text{Modified Words}}{\text{Total Words}} \times 100
\label{eq:perturbed_word}
\end{equation}
\item \textbf{Average Words per Input:} The average number of words per input sample across the dataset used for attack evaluation.
\item \textbf{Average Number of Queries:} The mean number of model inference queries made during each attack.
\item \textbf{Post-Attack Accuracy:} The classification accuracy of an ML model against the perturbed emails that were passed to the model after preprocessing defined in Eq. \ref{eq:post_accuracy}.
\begin{equation}
\small \textstyle \text{Post-Attack Accuracy (\%)} = \frac{\text{Correct Predictions on Perturbed Inputs}}{\text{Total Perturbed Inputs}} \times 100
\label{eq:post_accuracy}
\end{equation}
\end{itemize}

\subsection{Adversarial Attack Results}
Adversarial attacks were performed on approximately 10\% of the phishing emails within the Model Deployment evaluation set to assess the models' robustness against perturbed inputs. The successful adversarial phishing emails were recorded and evaluated again with the ML models trained with Baseline and Advanced datasets after preprocessing. As previously noted, the ML models trained with Word2Vec embeddings mostly outperformed those using TF-IDF and GloVe embeddings in both development and deployment settings. Therefore, for the sake of conciseness, we report adversarial evaluation results only for the models trained with Word2Vec embeddings. Table \ref{tab:deepwordbug_dataset1_allfeatures} and \ref{tab:deepwordbug_dataset2_allfeatures} display the results for adversarial attacks against Baseline dataset (Level 1 preprocessing) and Advanced dataset (Level 1 \& 2 preprocessing) for Word2Vec embedding.
\begin{table}[hbtp]
\centering
\caption{Attack results against models trained with Baseline dataset}
\label{tab:deepwordbug_dataset1_allfeatures}
\scriptsize
\begin{tabular}{|l|c|c|c|c|c|}
\hline
\textbf{Metric} & \textbf{LR} & \textbf{KNN} & \textbf{RF} & \textbf{MLP} & \textbf{SVM} \\
\hline
\multicolumn{6}{|c|}{\textbf{DeepWordBug}} \\
\hline
\# Successful Attacks & 423 & 89 & 251 & 318 & 409 \\
\# Failed Attacks & 15 & 327 & 166 & 142 & 11 \\
\# Skipped Attacks & 60 & 82 & 81 & 38 & 78 \\
Original Accuracy \% & 87.95 & 83.53 & 83.73 & \textbf{92.37} & 84.34 \\
Accuracy Under Attack \% & 3.01 & \textbf{65.66} & 33.33 & 28.51 & 2.21 \\
Attack Success Rate \% & 96.58 & 21.39 & 60.19 & 69.13 & 97.38 \\
Perturbed Word \% & 30.35 & 9.62 & 18.66 & 26.72 & 18.83 \\
Words Per Input & 31.03 & 31.03 & 31.03 & 31.03 & 42.80 \\
Queries & 76.46 & 137.42 & 109.31 & 100.19 & 71.39 \\
\hline
\multicolumn{6}{|c|}{\textbf{After Level 1 Preprocessing + Classification}} \\
\hline
Correctly Classified & 67 & 15 & 39 & 58 & 348 \\
Incorrectly Classified & 356 & 74 & 212 & 260 & 61 \\
Post-Attack Accuracy & 15.84 & 16.85 & 15.53 & 18.23 & \textbf{85.08} \\
\hline
\multicolumn{6}{|c|}{\textbf{Pruthi}} \\
\hline
\# Successful Attacks & 35 & 61 & 251 & 28 & 69 \\
\# Failed Attacks & 403 & 355 & 166 & 432 & 337 \\
\# Skipped Attacks & 60 & 82 & 81 & 38 & 92 \\
Original Accuracy \% & 87.95 & 83.53 & 83.73 & \textbf{92.37} & 81.53 \\
Accuracy Under Attack \% & 80.92 & 71.29 & 33.33 & \textbf{86.75} & 67.67 \\
Attack Success Rate \% & 7.99 & 14.66 & 60.19 & 6.09 & 17.0 \\
Perturbed Word \% & 5.90 & 5.98 & 18.66 & 6.62 & 5.72 \\
Words Per Input & 31.03 & 31.03 & 31.03 & 31.03 & 31.03 \\
Avg Queries & 1424.66 & 1413.71 & 109.31 & 1404.05 & 1397.99 \\
\hline
\multicolumn{6}{|c|}{\textbf{After Level 1 Preprocessing + Classification}} \\
\hline
Correctly Classified & 3 & 5 & 39 & 2 & 55 \\
Incorrectly Classified & 56 & 67 & 212 & 26 & 14 \\
Post-Attack Accuracy & 8.57 & 8.19 & 15.53 & 7.14 & \textbf{79.71} \\
\hline
\multicolumn{6}{|c|}{\textbf{TextBugger}} \\
\hline
Successful Attacks & 320 & 107 & 279 & 277 & 343 \\
Failed Attacks & 118 & 309 & 138 & 183 & 80 \\
Skipped Attacks & 60 & 82 & 81 & 38 & 75 \\
Original Accuracy \% & 87.95 & 83.53 & 83.73 & \textbf{92.37} & 84.94 \\
Accuracy Under Attack \% & 23.69 & \textbf{62.05} & 27.71 & 36.75 & 16.06 \\
Attack Success Rate \% & 73.06 & 25.72 & 66.91 & 60.22 & 81.09 \\
Perturbed Word \% & 58.30 & 25.77 & 52.35 & 52.94 & 57.41 \\
Words Per Input & 31.03 & 31.03 & 31.03 & 31.03 & 31.03 \\
Queries & 137.45 & 398.76 & 224.41 & 221.48 & 134.11 \\
\hline
\multicolumn{6}{|c|}{\textbf{After Level 1 Preprocessing + Classification}} \\
\hline
Correctly Classified & 115 & 18 & 91 & 115 & 123 \\
Incorrectly Classified & 205 & 89 & 188 & 162 & 220 \\
Post-Attack Accuracy & 35.94 & 16.82 & 32.62 & \textbf{41.52} & 35.86 \\
\hline
\multicolumn{6}{|c|}{\textbf{Custom}} \\
\hline
\# Successful Attacks & 365 & 88 & 258 & 276 & 337 \\
\# Failed Attacks & 73 & 328 & 159 & 184 & 122 \\
\# Skipped Attacks & 60 & 82 & 81 & 38 & 39 \\
Original Accuracy \% & 87.95 & 83.53 & 83.73 & \textbf{92.37} & 92.17 \\
Accuracy Under Attack \% & 14.66 & \textbf{65.86} & 31.93 & 36.95 & 24.50 \\
Attack Success Rate \% & 83.33 & 21.15 & 61.87 & 60.00 & 73.42 \\
Perturbed Word \% & 26.66 & 9.51 & 19.24 & 23.13 & 27.71 \\
Words Per Input & 31.03 & 31.03 & 31.03 & 31.03 & 31.03 \\
Queries & 145.91 & 325.48 & 242.32 & 212.30 & 143.82 \\
\hline
\multicolumn{6}{|c|}{\textbf{After Level 1 Preprocessing + Classification}} \\
\hline
Correctly Classified & 53 & 13 & 47 & 61 & 63 \\
Incorrectly Classified & 312 & 75 & 211 & 215 & 274 \\
Post-Attack Accuracy \% & 14.52 & 14.77 & 18.21 & \textbf{22.10} & 18.69 \\
\hline
\end{tabular}
\end{table}
\begin{table}[hbtp]
\centering
\caption{Attack results against models trained with Advanced dataset}
\label{tab:deepwordbug_dataset2_allfeatures}
\scriptsize
\begin{tabular}{|l|c|c|c|c|c|}
\hline
\textbf{Metric} & \textbf{LR} & \textbf{KNN} & \textbf{RF} & \textbf{MLP} & \textbf{SVM} \\
\hline
\multicolumn{6}{|c|}{\textbf{DeepWordBug}} \\
\hline
\# Successful Attacks & 431 & 90 & 181 & 412 & 411 \\
\# Failed Attacks & 12 & 149 & 135 & 23 & 26 \\
\# Skipped Attacks & 55 & 260 & 182 & 63 & 61 \\
Original Accuracy \% & \textbf{88.96} & 47.90 & 63.45 & 87.35 & 87.75 \\
Accuracy Under Attack \% & 2.41 & \textbf{29.86} & 27.11 & 4.62 & 5.22 \\
Attack Success Rate \% & 97.29 & 37.66 & 57.28 & 94.71 & 94.05 \\
Perturbed Word \% & 22.21 & 6.02 & 11.12 & 17.94 & 25.07 \\
Words Per Input & 42.80 & 42.73 & 42.80 & 42.80 & 42.80 \\
Queries & 78.27 & 141.44 & 123.27 & 73.03 & 82.21 \\
\hline
\multicolumn{6}{|c|}{\textbf{After Level 1 \& 2 Preprocessing + Classification}} \\
\hline
Correctly Classified & 406 & 49 & 158 & 366 & 338 \\
Incorrectly Classified & 25 & 56 & 23 & 46 & 98 \\
Post-Attack Accuracy & \textbf{94.19} & 46.67 & 87.29 & 88.83 & 77.52 \\
\hline
\multicolumn{6}{|c|}{\textbf{Pruthi}} \\
\hline
\# Successful Attacks & 25 & 72 & 181 & 51 & 34 \\
\# Failed Attacks & 418 & 164 & 135 & 384 & 403 \\
\# Skipped Attacks & 55 & 262 & 182 & 63 & 61 \\
Original Accuracy \% & \textbf{88.96} & 47.39 & 63.45 & 87.35 & 87.75 \\
Accuracy Under Attack \% & \textbf{83.94} & 32.93 & 27.11 & 77.11 & 80.92 \\
Attack Success Rate \% & 5.64 & 30.51 & 57.28 & 11.72 & 7.78 \\
Perturbed Word \% & 4.98 & 3.69 & 11.12 & 4.95 & 5.66 \\
Words Per Input & 42.80 & 42.80 & 42.80 & 42.80 & 42.80 \\
Queries & 1140.74 & 1155.32 & 123.27  & 1107.89 & 1132.68 \\
\hline
\multicolumn{6}{|c|}{\textbf{After Level 1 \& 2 Preprocessing + Classification}} \\
\hline
Correctly Classified & 14 & 34 & 158 & 27 & 24 \\
Incorrectly Classified & 11 & 38 & 23 & 24 & 10 \\
Post-Attack Accuracy \% & 71.79 & 47.22 & \textbf{87.29} & 52.94 & 70.58 \\
\hline
\multicolumn{6}{|c|}{\textbf{TextBugger}} \\
\hline
Successful Attacks & 415 & 128 & 263 & 408 & 335 \\
Failed Attacks & 28 & 108  & 53 & 27 & 119 \\
Skipped Attacks & 55 & 262 & 182 & 63 & 44 \\
Original Accuracy & 88.96 & 47.39 & 63.45 & 87.35 & \textbf{91.16} \\
Accuracy Under Attack & 5.62 & 21.69 & 10.64 & 5.42 & \textbf{23.90} \\
Attack Success Rate  & 93.68 & 54.24 & 83.23 & 93.79 & 73.79 \\
Perturbed Word \%  & 57.27 & 22.01 & 37.16 & 54.41 & 67.69 \\
Words Per Input & 42.80 & 42.80 & 42.80 & 42.80 & 42.80 \\
Queries & 155.15 & 326.19  & 207.61 & 137.76 & 182.67 \\
\hline
\multicolumn{6}{|c|}{\textbf{After Level 1 \& 2 Preprocessing + Classification}} \\
\hline
Correctly Classified & 232 & 50  & 205 & 257 & 227 \\
Incorrectly Classified & 183 & 78 & 58 & 151 & 108 \\
Post-Attack Accuracy & 55.90 & 39.06 & \textbf{77.94} & 62.99 & 67.76 \\
\hline
\multicolumn{6}{|c|}{\textbf{Custom}} \\
\hline
\# Successful Attacks & 374 & 104 & 182 & 382 & 381 \\
\# Failed Attacks & 69 & 132 & 134 & 53 & 39 \\
\# Skipped Attacks & 55 & 262 & 182 & 63 & 78 \\
Original Accuracy \% & \textbf{88.96} & 47.39 & 63.45 & 87.35 & 84.34 \\
Accuracy Under Attack \% & 13.86 & 26.51 & \textbf{26.91} & 10.64 & 7.83 \\
Attack Success Rate \% & 84.42 & 44.07 & 57.59 & 87.82 & 90.71 \\
Perturbed Word \% & 19.56 & 5.42 & 11.18 & 16.37 & 17.36 \\
Words Per Input & 42.80 & 42.80 & 42.80 & 42.80 & 42.80 \\
Queries & 121.55 & 258.81 & 236.86 & 113.68 & 107.57 \\
\hline
\multicolumn{6}{|c|}{\textbf{After Level 1 \& 2 Preprocessing + Classification}} \\
\hline
Correctly Classified & 350 & 46 & 161 & 340 & 331 \\
Incorrectly Classified & 24 & 58 & 21 & 42 & 50 \\
Post-Attack Accuracy \% & \textbf{93.58} & 44.23 & 88.46 & 89.00 & 86.87 \\
\hline
\end{tabular}
\end{table}
\subsubsection{DeepWordBug}
For Level 1 preprocessing, i.e., models developed with Baseline dataset, the MLP based model achieved the highest original accuracy of 92.37\%. KNN based model achieved highest accuracy under attack (65.66\%). The highest post-attack accuracy was achieved by SVM with accuracy of 85.08\%. In terms of susceptibility, SVM had the highest attack success rates of 97.38\%, indicating that the model is the most vulnerable. 

For Level 1 and 2 preprocessing, i.e., models developed with Advanced dataset, the best original accuracy was achieved by LR (88.96\%). KNN based model again achieved highest accuracy under attack (29.86\%). In terms of post-attack accuracy, the highest performance was observed for LR (94.19\%) making it the most resilient configuration. The worst post-attack accuracy was recorded for KNN (46.67\%), which is significantly higher than LR, KNN, RF, and MLP based models for Level 1 preprocessing. All of them were achieving accuracy of less than 20\%.   

\subsubsection{Pruthi Attack}
For Level 1 preprocessing, MLP based model achieved highest accuracy under attack (86.75\%). The highest post-attack accuracy was achieved by SVM with an accuracy of 79.71\%. RF was the most susceptible model with the attack success rate of 60.19\%. 

The addition of Level 2 preprocessing resulted in a substantial improvement in robustness across all models except the SVM. LR based model achieved highest accuracy under attack (83.94\%). In terms of post-attack accuracy, the highest performance was observed for RF (87.29\%) making it the most resilient configuration. The worst post-attack accuracy was recorded for KNN (47.22\%), which is significantly higher than LR, KNN, RF, and MLP based models for Level 1 preprocessing. All of them were achieving accuracy of less than 16\%. 
\subsubsection{TextBugger Attack}
For Level 1 preprocessing, KNN based model achieved highest accuracy under attack (62.05\%). The highest post-attack accuracy was achieved by MLP with an accuracy of 41.52\%. SVM was the most susceptible model with the attack success rate of 81.09\%.

Combining Level 1 and 2 preprocessing resulted in improvement in robustness across all the models. SVM based model achieved highest accuracy under attack (23.90\%). In terms of post-attack accuracy, the highest performance was observed for RF (77.94\%) making it the most resilient configuration. The worst post-attack accuracy was recorded for KNN (39.06\%), which is comparable to the best performing MLP based model for Level 1 preprocessing. 
\subsubsection{Custom Attack}
For Level 1 preprocessing, KNN based model achieved highest accuracy under attack (65.86\%). The highest post-attack accuracy was achieved by MLP with an accuracy of 22.10\%. LR was the most susceptible model with the attack success rate of 83.33\%. 

For Level 1 and 2 preprocessing, RF based model achieved highest accuracy under attack (26.91\%). In terms of post-attack accuracy, the highest performance was observed for LR (93.58\%) making it the most resilient configuration. The worst post-attack accuracy was recorded for KNN (44.23\%), which is significantly higher than all the models for Level 1 preprocessing. All of them were achieving accuracy of less than 23\%.   

\begin{figure}[h!]
\centering
\includegraphics[width=0.95\columnwidth]{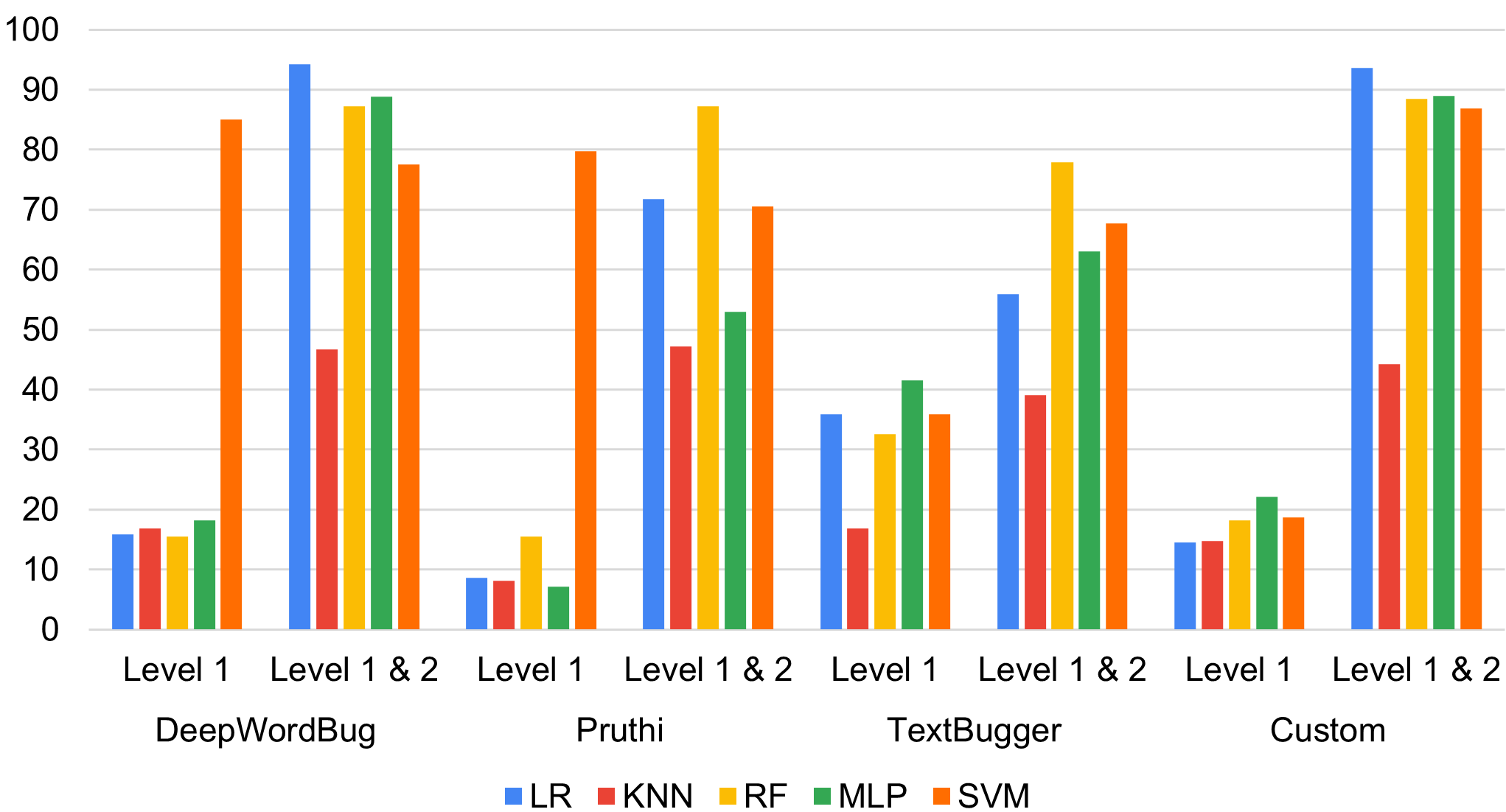}
\caption{Comparison of post-attack accuracy for models trained with Word2Vec embedding}
\label{fig:post_attack_accuracy}
\end{figure}

\subsection{Summary of Adversarial Attacks}
The experimental results for all four adversarial attacks—DeepWordBug, Pruthi, TextBugger, and the Custom Attack—demonstrate that the vulnerability of phishing detection models depends on both the classifier architecture and the preprocessing techniques employed. The combination of Level 1 and 2 preprocessing used in Advanced dataset significantly mitigated the impact of adversarial attacks. Models that initially performed poorly under attack conditions showed improvements in post-attack accuracy, indicating that robust preprocessing pipelines are critical for the deployment of phishing detection systems in real-world scenarios. Figure \ref{fig:post_attack_accuracy} illustrates the comparison of post-attack accuracies across classifiers using Word2Vec embeddings. In summary, these findings support our \textbf{RQ2}, confirming that the proposed Level 1 and Level 2 preprocessing techniques substantially enhance the adversarial robustness of phishing email detectors.
\section{Performance on LLM Generated Phishing Emails}
\label{sec:llm_based_phishing}
To further evaluate model robustness, phishing emails were synthetically generated using LLMs including ChatGPT and Llama. Prompt engineering principles were employed to ensure the generated samples were realistic, varied, and challenging for the classifier. The prompt construction begins with an explicit role instruction - \textbf{“act as a phishing email generator”} - which contextualizes the model’s behavior and aligns its output with the task objective. Role-based prompting increases the likelihood of generating relevant and coherent content. The model was instructed to impersonate trusted brands in the USA, apply social engineering techniques, include misleading URLs, and align subject lines with message content. Constraints were also applied to ensure demographic relevance by targeting audiences in the USA and to enforce a realistic message length. To guide the model’s generation capabilities, examples were provided as part of a few-shot learning setup. This primes the LLM with recognizable patterns and reduces output randomness. The output format was predefined as a CSV with structured fields: \texttt{Sender}, \texttt{Subject}, and \texttt{Content}, allowing for easy integration with phishing detection pipelines. Duplication was discouraged in the subject lines, and the model was asked to produce 15 unique phishing samples in a single generation pass, facilitating efficient batch creation across models. The ChatGPT GPT-4o (unpaid subscription) was used, while Llama 3 was deployed locally using the \texttt{Ollama} \cite{oLlama-download} runtime. 

Despite carefully crafted prompts, the generation of phishing emails using LLMs faced  challenges due to inherent guardrails embedded within these models \cite{dong2024safeguarding}. These guardrails, implemented to uphold ethical use and prevent malicious generation, limited the LLM's willingness to produce realistic phishing content. As a result, while approximately 300 phishing emails were initially generated across multiple runs, a significant portion of them were failed to convincingly mimic phishing attempts. After manual inspection based on quality criteria such as message realism and adherence to prompts only 88 phishing emails were selected for final evaluation. Among them, 43 and 45 were generated by Chatgpt and Llama, respectively. This elimination process ensured that only samples representative of true phishing threats were tested, maintaining the integrity and robustness of the model evaluation pipeline.
\begin{table}[h!]
\centering
\caption{Performance on LLM generated phishing emails}
\label{tab:llm_results}
\begin{tabular}{|l|c|c|c|c|}
\hline
\textbf{Feature Type} & \textbf{Model} & \textbf{Level 1 Preprocessing (\%)} & \textbf{Level 1 \& 2 Preprocessing (\%)} \\
\hline
\multirow{5}{*}{\textbf{TF-IDF}} 
& \cellcolor{lightgray} LR & \cellcolor{lightgray} 67.04 & \cellcolor{lightgray} 72.72 \\
& SVM & 79.54 & 76.13 \\
& \cellcolor{lightgray} RF & \cellcolor{lightgray} 77.27 & \cellcolor{lightgray} 78.41 \\
& KNN & 85.22 & 84.09 \\
& \cellcolor{lightgray} MLP & \cellcolor{lightgray} 75.00 & \cellcolor{lightgray} 77.27 \\
\hline
\multirow{5}{*}{\textbf{Word2Vec}} 
& LR & 97.73 & 92.05 \\
& SVM & 100.00 & 92.05 \\
& RF & 95.45 & 89.77 \\
& KNN & 93.18 & 46.59 \\
& \cellcolor{lightgray} MLP & \cellcolor{lightgray} 97.72 & \cellcolor{lightgray} 100.00 \\
\hline
{\multirow{5}{*}{\textbf{GloVe}}} 
& \cellcolor{lightgray} LR & \cellcolor{lightgray} 85.27 & \cellcolor{lightgray} 88.63 \\
& \cellcolor{lightgray} SVM & \cellcolor{lightgray} 92.04 & \cellcolor{lightgray} 93.18 \\
& \cellcolor{lightgray} RF & \cellcolor{lightgray} 93.18 & \cellcolor{lightgray}97.72 \\
& \cellcolor{lightgray} KNN & \cellcolor{lightgray} 70.45 & \cellcolor{lightgray} 85.22 \\
& \cellcolor{lightgray} MLP & \cellcolor{lightgray} 86.36 & \cellcolor{lightgray} 89.77 \\
\hline
\end{tabular}
\end{table}

As shown in Table \ref{tab:llm_results}, the MLP model with Word2Vec embeddings achieves the highest performance, reaching 100.00\% accuracy with both Level 1 and 2 preprocessing. In contrast, LR with TF-IDF features yields the lowest accuracy, particularly under Level 1 preprocessing, with only 67.04\%. In most cases, the application of Level 1 and 2 preprocessing significantly enhances model performance—for example, KNN with GloVe improves from 70.45\% to 85.22\%, as indicated by the light gray-highlighted rows. These findings support our \textbf{RQ3}, demonstrating the effectiveness of Level 1 and 2 preprocessing in detecting phishing emails generated by LLMs. Table \ref{tab:comparison} provides a brief comparison of our work with the existing works. The performance reported in the table is based on model development settings.
\section{Conclusion and Future Work}
\label{sec:conclusion}
In this work, we proposed a robust phishing email detection system by integrating advanced preprocessing techniques and evaluating performance across multiple ML models and feature extraction methods. The inclusion of additional preprocessing that combined spelling correction and word splitting enhanced the detection for most of the ML models during model development and deployment stages. The MLP based model achieved the peak F1-scores of 99.49\% and 84.39\%, respectively for development and deployment with Level 1 and 2 preprocessing. Adversarial testing using attacks including TextBugger, DeepWordBug, Pruthi, and custom perturbations validated the adversarial robustness of phishing detection models. The results demonstrated that models trained with additional preprocessing were consistently more resilient to adversarial inputs. Additionally, the system exhibited improved detection capabilities against LLM generated phishing emails, showcasing its practical applicability in modern threat scenarios. 

As a part of the future work, we will build phishing detection system by finetuning open source LLMs such as BERT (with variants) and Llama to capture deeper context and meaning in phishing attempts. Detection can also be extended by including features from email headers like sender domain, IP traces alongside the email body. To enhance user trust in the phishing detection model, future work will focus on classifying phishing emails into specific categories (e.g., job scams, fake login attempts) using multi-class classifiers and providing explanations through explainable AI techniques. 
\begin{landscape}
\begin{table*}[htbp]
\centering
\footnotesize
\setlength{\tabcolsep}{2pt}
\caption{Comparison of our work with the existing works\\{\scriptsize (A: Accuracy, P: Precision, R: Recall, F: F1-score, TF: TF-IDF, and W2V: Word2Vec)}}
\label{tab:comparison}
\begin{tabular}{|c|c|c|c|c|c|c|c|c|}
\hline
\textbf{Paper} & \makecell{\textbf{Dataset Used}\\(Enron, Nazario, Millersmile)} & \makecell{\textbf{Timeline}} & \makecell{\textbf{Attack}} & \makecell{\textbf{Preprocessing}\\\textbf{(Correction, Splitting)}} & \makecell{\textbf{LLM}\\\textbf{Emails}} & \makecell{\textbf{Evaluation}\\\textbf{Metrics}\\(A, P, R, F)} & \makecell{\textbf{Feature Extractor}\\(TF, W2V, GloVe)} & \makecell{\textbf{Models}\\(LR, RF, SVM, \\ MLP, KNN)} \\
\hline
\cite{unnithan2018phishing}  & \xmark, \xmark, \xmark & \xmark & \xmark & \xmark, \xmark & \xmark & 88.4,--,--,-- & \cmark, \xmark, \xmark & \cmark, \cmark, \cmark, \xmark, \cmark \\
\hline
\cite{harikrishnan2018machine} & \xmark, \xmark, \xmark & \xmark & \xmark & \xmark, \xmark & \xmark & --, --, --, 94.4 & \cmark, \xmark, \xmark & \cmark, \cmark, \cmark, \xmark, \cmark \\
\hline
\cite{bountakas2021comparison} & \cmark, \cmark, \xmark & \xmark & \xmark & \xmark, \xmark & \xmark & 98.95, 98.63, 99.31, 98.97 & \cmark, \cmark, \xmark & \cmark, \cmark, \xmark, \xmark, \xmark \\
\hline
\cite{bountakas2023helphed} & \cmark, \cmark, \xmark & \cmark & \xmark & \xmark, \xmark & \xmark & 99.43, --, --, 99.42 & \xmark, \cmark, \xmark & \cmark, \cmark, \cmark, \xmark, \xmark \\
\hline
\cite{omari2023comparative} & \xmark, \xmark, \xmark & \xmark & \xmark & \xmark, \xmark & \xmark & 97.2, 96.8, 97.0, 96.9 & \xmark, \xmark, \xmark & \cmark, \cmark, \cmark, \xmark, \cmark \\
\hline
\cite{somesha2024deepphishnet} & \xmark, \xmark, \xmark & \xmark & \xmark & \xmark, \xmark & \xmark & 99.52, --, --, -- & \cmark, \cmark, \xmark & \xmark, \xmark, \xmark, \xmark, \xmark \\
\hline
\cite{sentence_knn_2022} & \xmark, \xmark, \xmark & \xmark & \xmark & \xmark, \xmark & \xmark & 97.0, --, --, -- & \xmark, \cmark, \xmark & \xmark, \xmark, \xmark, \xmark, \cmark \\
\hline
\cite{syntactic_features_2015} & \cmark, \cmark, \xmark & \xmark & \xmark & \xmark, \xmark & \xmark & --, --, --, -- & \xmark, \xmark, \xmark & \xmark, \xmark, \xmark, \xmark, \xmark \\
\hline
\cite{gong2019context} & \xmark, \xmark, \xmark & \xmark & \xmark & \xmark, \cmark & \xmark & --, 72.1, 75.7, 73.7 & \xmark, \cmark, \xmark & \xmark, \xmark, \xmark, \xmark, \xmark \\
\hline
\cite{feature_perturbation_2022} & \xmark, \cmark, \xmark & \xmark & \cmark & \xmark, \xmark & \xmark & 99.83, --, --, -- & \cmark, \cmark, \xmark & \cmark, \xmark, \cmark, \cmark, \xmark \\
\hline
\cite{agwep_2023} & \xmark, \xmark, \xmark & \xmark & \xmark & \xmark, \xmark & \xmark & --, 96.0, 92.0, -- & \xmark, \xmark, \cmark & \xmark, \xmark, \xmark, \xmark, \xmark \\
\hline
\cite{robustness_phishing_2023} & \xmark, \xmark, \xmark & \xmark & \cmark & \xmark, \xmark & \xmark & 99.0, --, --, 95.0 & \xmark, \xmark, \cmark & \xmark, \xmark, \xmark, \xmark, \xmark \\
\hline
\cite{bert_adversarial_2023} & \xmark, \xmark, \xmark & \xmark & \cmark & \xmark, \xmark & \xmark & 98.39, 95.12, 93.98, 94.55 & \xmark, \xmark, \xmark & \xmark, \xmark, \xmark, \xmark, \xmark \\
\hline
\cite{phishbots_llms_2023} & \cmark, \cmark, \xmark & \xmark & \xmark & \xmark, \xmark & \cmark & 94, --, --, -- & \xmark, \xmark, \xmark & \xmark, \xmark, \xmark, \xmark, \xmark \\
\hline
\cite{chatgpt_spam_2023} & \cmark, \xmark, \xmark & \xmark & \xmark & \xmark, \xmark & \cmark & 99.4, 99.7, 98.8, -- & \xmark, \xmark, \xmark & \xmark, \xmark, \xmark, \xmark, \xmark \\
\hline
\textbf{Our work} & \cmark, \cmark, \cmark & \cmark & \cmark & \cmark,\cmark & \cmark & 99.99, 99.62, 99.38, 99.49 & \cmark, \cmark, \cmark & \cmark, \cmark, \cmark, \cmark, \cmark \\
\hline
\end{tabular}
\end{table*}
\end{landscape}

\bibliographystyle{IEEEtran}
\bibliography{references}

\begin{thebibliography}{10}
\providecommand{\url}[1]{#1}
\csname url@samestyle\endcsname
\providecommand{\newblock}{\relax}
\providecommand{\bibinfo}[2]{#2}
\providecommand{\BIBentrySTDinterwordspacing}{\spaceskip=0pt\relax}
\providecommand{\BIBentryALTinterwordstretchfactor}{4}
\providecommand{\BIBentryALTinterwordspacing}{\spaceskip=\fontdimen2\font plus
\BIBentryALTinterwordstretchfactor\fontdimen3\font minus \fontdimen4\font\relax}
\providecommand{\BIBforeignlanguage}[2]{{%
\expandafter\ifx\csname l@#1\endcsname\relax
\typeout{** WARNING: IEEEtran.bst: No hyphenation pattern has been}%
\typeout{** loaded for the language `#1'. Using the pattern for}%
\typeout{** the default language instead.}%
\else
\language=\csname l@#1\endcsname
\fi
#2}}
\providecommand{\BIBdecl}{\relax}
\BIBdecl

\bibitem{gallup2024survey}
L.~Saad, ``{Scams: Relatively Common and Anxiety-Inducing for Americans},'' 2023, https://news.gallup.com/poll/544643/scams-relatively-common-anxiety-inducing-americans.aspx.

\bibitem{gallup2024demo}
L.~DeNicola, ``{Who Gets Scammed the Most?}'' 2024, https://www.experian.com/blogs/ask-experian/who-gets-scammed-most/.

\bibitem{economicimpact2024}
FBI, ``{FBI Releases the Internet Crime Complaint Center 2018 Internet Crime Report},'' 2019, https://www.fbi.gov/news/press-releases/fbi-releases-the-internet-crime-complaint-center-2018-internet-crime-report.

\bibitem{fraudgpt}
Z.~Amos, ``{What Is FraudGPT?}'' https://hackernoon.com/what-is-fraudgpt, 2023.

\bibitem{wormgpt}
D.~Kelley, ``{WormGPT – The Generative AI Tool Cybercriminals Are Using to Launch Business Email Compromise Attacks},'' https://slashnext.com/blog/wormgpt-the-generative-ai-tool-cybercriminals-are-using-to-launch-business-email-compromise-attacks/, 2023.

\bibitem{llmimpact2024}
B.~S. Fredrik~Heiding and A.~Vishwanath, ``{AI Will Increase the Quantity — and Quality — of Phishing Scams},'' 2024, https://hbr.org/2024/05/ai-will-increase-the-quantity-and-quality-of-phishing-scams.

\bibitem{llmsocialengineering2024}
D.~Manky and G.~Baram, ``Beyond phishing: Exploring the rise of ai-enabled cybercrime,'' 2025, https://cltc.berkeley.edu/2025/01/16/beyond-phishing-exploring-the-rise-of-ai-enabled-cybercrime/.

\bibitem{textattack}
J.~X. Morris, E.~Lifland, J.~Y. Yoo, J.~Grigsby, D.~Jin, and Y.~Qi, ``Textattack: A framework for adversarial attacks, data augmentation, and adversarial training in nlp,'' \emph{arXiv preprint arXiv:2005.05909}, 2020.

\bibitem{chatgpt}
``{ChatGPT},'' \url{https://chatgpt.com/}.

\bibitem{llama}
``{Llama},'' \url{https://www.llama.com/}.

\bibitem{kulal}
D.~Kulal, L.~Shiferaw, and Q.~Niyaz, ``Phishing email detection through machine learning and word error correction,'' in \emph{2025 17th International Conference on COMmunication Systems and NETworks (COMSNETS)}, 2025, pp. 1299--1304.

\bibitem{unnithan2018phishing}
S.~Unnithan, R.~Vinayakumar, and P.~Poornachandran, ``Phishing email detection using machine learning techniques,'' in \emph{International Conference on Advances in Computing, Communications and Informatics (ICACCI)}.\hskip 1em plus 0.5em minus 0.4em\relax IEEE, 2018.

\bibitem{harikrishnan2018machine}
N.~Harikrishnan, R.~Vinayakumar, and K.~Soman, ``A machine learning approach towards phishing email detection,'' in \emph{Proceedings of the anti-phishing pilot at ACM international workshop on security and privacy analytics (IWSPA AP)}, vol. 2013, 2018, pp. 455--468.

\bibitem{bountakas2021comparison}
P.~Bountakas, K.~Koutroumpouchos, and C.~Xenakis, ``A comparison of natural language processing and machine learning methods for phishing email detection,'' in \emph{Proceedings of the 16th International Conference on Availability, Reliability and Security}, 2021, pp. 1--12.

\bibitem{bountakas2023helphed}
P.~Bountakas and C.~Xenakis, ``Helphed: Hybrid ensemble learning phishing email detection,'' \emph{Journal of network and computer applications}, vol. 210, p. 103545, 2023.

\bibitem{omari2023comparative}
K.~Omari, ``Comparative study of machine learning algorithms for phishing website detection,'' \emph{International Journal of Advanced Computer Science and Applications}, vol.~14, no.~9, 2023.

\bibitem{somesha2024deepphishnet}
G.~Somesha and A.~Pais, ``Deepphishnet: A deep learning framework for phishing email detection using word embeddings,'' \emph{Journal of Cybersecurity and Privacy}, vol.~4, no.~1, pp. 15--34, 2024.

\bibitem{sentence_knn_2022}
L.~Sawe, J.~Gikandi, J.~Kamau, and D.~Njuguna, ``Sentence level analysis model for phishing detection using knn,'' \emph{Journal of Cybersecurity}, vol.~6, pp. 2579--0072, 2024.

\bibitem{syntactic_features_2015}
G.~Park and J.~M. Taylor, ``Using syntactic features for phishing detection,'' \emph{arXiv preprint arXiv:1506.00037}, 2015.

\bibitem{gong2019context}
H.~Gong, Y.~Li, S.~Bhat, and P.~Viswanath, ``Context-sensitive malicious spelling error correction,'' in \emph{The World Wide Web Conference}, 2019, pp. 2771--2777.

\bibitem{feature_perturbation_2022}
Q.~Cheng, A.~Xu, X.~Li, and L.~Ding, ``Adversarial email generation against spam detection models through feature perturbation,'' in \emph{2022 IEEE International Conference on Assured Autonomy (ICAA)}.\hskip 1em plus 0.5em minus 0.4em\relax IEEE, 2022, pp. 83--92.

\bibitem{agwep_2023}
J.~Gregory and Q.~Liao, ``Adversarial spam generation using adaptive gradient-based word embedding perturbations,'' in \emph{2023 IEEE International Conference on Artificial Intelligence, Blockchain, and Internet of Things (AIBThings)}.\hskip 1em plus 0.5em minus 0.4em\relax IEEE, 2023, pp. 1--5.

\bibitem{robustness_phishing_2023}
P.~Mehdi~Gholampour and R.~M. Verma, ``Adversarial robustness of phishing email detection models,'' in \emph{Proceedings of the 9th ACM international workshop on security and privacy analytics}, 2023, pp. 67--76.

\bibitem{bert_adversarial_2023}
A.~Kushwaha, K.~Dutta, and V.~Maheshwari, ``Analysis of bert email spam classifier against adversarial attacks,'' in \emph{2023 International Conference on Artificial Intelligence and Smart Communication (AISC)}.\hskip 1em plus 0.5em minus 0.4em\relax IEEE, 2023, pp. 485--490.

\bibitem{phishbots_llms_2023}
S.~S. Roy, P.~Thota, K.~V. Naragam, and S.~Nilizadeh, ``From chatbots to phishbots?--preventing phishing scams created using chatgpt, google bard and claude,'' \emph{arXiv preprint arXiv:2310.19181}, 2023.

\bibitem{chatgpt_spam_2023}
A.~Utaliyeva, M.~Pratiwi, H.~Park, and Y.-H. Choi, ``Chatgpt: A threat to spam filtering systems,'' in \emph{2023 IEEE International Conference on High Performance Computing \& Communications, Data Science \& Systems, Smart City \& Dependability in Sensor, Cloud \& Big Data Systems \& Application (HPCC/DSS/SmartCity/DependSys)}.\hskip 1em plus 0.5em minus 0.4em\relax IEEE, 2023, pp. 1043--1050.

\bibitem{millersmile}
Millersmiles, ``millersmiles.co.uk,'' \url{http://www.millersmiles.co.uk/links .php}, 2016.

\bibitem{Nazario}
J.~Nazario, ``Zenodo-phishing dataset,'' \url{https://zenodo.org/record/8339691}, 2016.

\bibitem{Enron}
W.~Kirk, ``The enron email dataset,'' \url{https://www.kaggle.com/datasets/wcukierski/enron-email-dataset}, 2015.

\bibitem{langdetect}
``Langdetect,'' \url{https://pypi.org/project/langdetect/}, 2021.

\bibitem{googletrans}
``Googletrans,'' \url{https://pypi.org/project/googletrans/}, 2025.

\bibitem{pyspellchecker}
``Pyspellchecker,'' \url{https://pypi.org/project/pyspellchecker/}, 2024.

\bibitem{autocorrect}
``Autocorrect,'' \url{https://pypi.org/project/autocorrect/}, 2021.

\bibitem{textblob}
``Textblob,'' \url{https://pypi.org/project/textblob/}, 2025.

\bibitem{symspellpy}
``Symspellpy,'' \url{https://pypi.org/project/symspellpy/}, 2025.

\bibitem{wordninja}
Python, ``wordninja 2.0.0,'' \url{https://pypi.org/project/wordninja/}, 2019.

\bibitem{wordsegment}
``Wordsegment,'' \url{https://pypi.org/project/wordsegment/}, 2018.

\bibitem{spacy}
``Spacy,'' \url{https://pypi.org/project/spacy/}, 2025.

\bibitem{tfidf}
K.~Sparck~Jones, ``A statistical interpretation of term specificity and its application in retrieval,'' \emph{Journal of Documentation}, vol.~28, no.~1, pp. 11--21, 1972.

\bibitem{word2vec}
T.~Mikolov, K.~Chen, G.~Corrado, and J.~Dean, ``Efficient estimation of word representations in vector space,'' \emph{arXiv preprint arXiv:1301.3781}, 2013.

\bibitem{gensim}
``{Gensim},'' \url{https://pypi.org/project/gensim/}.

\bibitem{glove}
J.~Pennington, R.~Socher, and C.~D. Manning, ``Glove: Global vectors for word representation,'' in \emph{Proceedings of the 2014 conference on empirical methods in natural language processing (EMNLP)}, 2014, pp. 1532--1543.

\bibitem{glovefile}
C.~D.~M. Jeffrey~Pennington, Richard~Socher, ``Glove: Global vectors for word representation,'' \url{https://nlp.stanford.edu/projects/glove/}, 2014.

\bibitem{scikit-learn}
{scikit-learn developers}, ``Scikit-learn: Machine learning in python,'' \url{https://scikit-learn.org/stable/}, 2025, accessed: 2025-04-11.

\bibitem{gao2018deepwordbug}
J.~Gao, J.~Lanchantin, M.~L. Soffa, and Y.~Qi, ``Black-box generation of adversarial text sequences to evade deep learning classifiers,'' \emph{2018 IEEE Security and Privacy Workshops (SPW)}, 2018.

\bibitem{pruthi2019combating}
D.~Pruthi, B.~Dhingra, and Z.~C. Lipton, ``Combating adversarial misspellings with robust word recognition,'' in \emph{Proceedings of ACL}, 2019, pp. 5582--5591.

\bibitem{li2018textbugger}
J.~Li, S.~Ji, T.~Du, B.~Li, and T.~Wang, ``Textbugger: Generating adversarial text against real-world applications,'' \emph{Network and Distributed Systems Security (NDSS) Symposium}, 2019.

\bibitem{oLlama-download}
``Ollama,'' \url{https://ollama.com/}, 2025, accessed: 2025-04-11.

\bibitem{dong2024safeguarding}
Y.~Dong, R.~Mu, Y.~Zhang, S.~Sun, T.~Zhang, C.~Wu, G.~Jin, Y.~Qi, J.~Hu, J.~Meng \emph{et~al.}, ``Safeguarding large language models: A survey,'' \emph{arXiv preprint arXiv:2406.02622}, 2024.

\end{thebibliography}

\end{document}